\documentclass[11pt,a4paper]{article}
\usepackage{graphicx}
\usepackage{latexsym}
\usepackage{subfigure}
\usepackage{amsmath}
\usepackage{enumerate}
\usepackage{amssymb}
\usepackage[mathscr]{eucal}
\usepackage{amsthm}
\usepackage{color}
\usepackage[small]{caption}
\usepackage[hypertexnames=false,colorlinks=true,linkcolor=blue,citecolor=blue,
            bookmarksopen=false,bookmarks=false,
            pdfstartview=XYZ,pdffitwindow=true,pdfcenterwindow=true]{hyperref}
\usepackage[a4paper,text={6.0in,9.0in},centering,
            includefoot,foot=0.6in]{geometry}
\usepackage[printfigures]{figcaps}
\usepackage{authblk}
\allowdisplaybreaks[1]

\newtheorem{thm}{Theorem}[section]

\newtheorem{lem}[thm]{Lemma}

\theoremstyle{definition}
\newtheorem{ex}[thm]{Example}

\theoremstyle{remark}

\renewcommand{\d}{\mathrm{d}}

\begin{document}

\figcapsoff

\title{Patch dynamics with buffers for homogenization problems}%
\author[1]{Giovanni Samaey} 
\author[1]{Dirk Roose}
\author[2]{Ioannis G. Kevrekidis}
\affil[1]{\small{
Dept. of Computer Science, K.U. Leuven,
Celestijnenlaan 200A, 3001 Leuven, Belgium}}
\affil[2]{\small{
Dept. of Chemical Engineering and PACM,
Princeton University, Princeton, NJ08544}}

\date{\today}

\maketitle

\begin{abstract}
An important class of problems exhibits smooth behaviour on macroscopic space
and time scales, while only a microscopic evolution law is known.  For such
time-dependent multi-scale problems, an ``equation-free" framework has been
proposed, of which patch dynamics is an essential component.  Patch dynamics is
designed to perform numerical simulations of an unavailable macroscopic
equation on macroscopic time and length scales; it uses appropriately
initialized simulations of the available microscopic model in a number of small
boxes (patches), which cover only a fraction of the space-time domain.  To
reduce the effect of the artificially introduced box boundaries, we use buffer
regions to ``shield" the boundary artefacts from the interior of the domain for
short time intervals.  We analyze the accuracy of this scheme for a 
diffusion homogenization problem with periodic heterogeneity, and propose a
simple heuristic to determine a sufficient buffer size.  The algorithm
performance is illustrated through a set of numerical examples, which include
a non-linear reaction-diffusion equation and the Kuramoto--Sivashinsky
equation.  \end{abstract}

\clearpage

\section{Introduction \label{sec:introduction}} 

For an important class of multi-scale problems, a separation of scales prevails
between the (microscopic, detailed) level of description of the available
model, and the (macroscopic, continuum) level at which one would like to
observe and analyze the system.  Consider, for example, a kinetic Monte Carlo
model of bacterial growth \cite{SetGearOthKevr03}.  A stochastic model
describes the probability of an individual bacterium to run or ``tumble", based
on the rotation of its flagellae.  Technically, it would be possible to simply
evolve the detailed model and observe the macroscopic variables of interest
(e.g.~cell density), but this could be prohibitively expensive.  It is known,
however, that, under certain conditions, one could write a deterministic
equation for the evolution of the macroscopic observable (here \emph{bacteria
concentration}, the zeroth moment of the evolving distribution) on macroscopic
space and time scales, but it is hard to obtain an accurate closed formula
explicitly.

The recently proposed \emph{equation-free framework}
\cite{KevrGearHymKevrRunTheo02} can then be used instead of stochastic time
integration in the entire space-time domain.  This framework is built around
the central idea of a \emph{coarse time-stepper}, which is a time-$\delta t$
map from coarse variables to coarse variables.  It consists of the following
steps: (1) {\it lifting}, i.e.~the creation of \emph{appropriate} initial
conditions for the microscopic model; (2) {\it evolution}, using the
microscopic model and (possibly) some constraints; and (3) {\it restriction},
i.e.~the projection of the detailed solution to the macroscopic observation
variables.  This coarse time-stepper can subsequently be used as ``input" for
time-stepper based algorithms performing macroscopic numerical analysis tasks.
These include, for example, time-stepper based bifurcation codes to perform
bifurcation analysis for the \emph{unavailable} macroscopic equation
\cite{MakMarKevr02,MakMarPanKevr02,TheoQianKevr00,TheoSanSunKevr02}.  This
approach has already been used in several applications
\cite{HumKevr02,SietGrahKevr02}, and also allows to perform other system level
tasks, such as control and optimization \cite{SietArmMakKevr02}.

When dealing with systems that would be described by (in our case, unavailable)
\emph{partial} differential equations (PDEs), one can also reduce the {\it
spatial} complexity.  For systems with one space dimension, the \emph{gap-tooth
scheme} \cite{KevrGearHymKevrRunTheo02} was proposed; it can be generalized in
several space dimensions.  A number of small intervals, separated by large
gaps, are introduced; they qualitatively correspond to mesh points for a
traditional, continuum solution of the unavailable equation.  In higher space
dimensions, these intervals would become \emph{boxes} around the coarse mesh
points, a term that we will also use throughout this paper.  We construct a
coarse time-$\delta t$ map as follows.  We first choose a number of macroscopic
grid points.  Then, we choose a small interval around each grid point;
initialize the fine scale, microscopic solver within each interval consistently
with the macroscopic initial condition profiles; and provide each box with
appropriate boundary conditions.  Subsequently, we use the microscopic model in
each interval to simulate until time $\delta t$, and obtain macroscopic
information (e.g.\ by computing the average density in each box) at time
$\delta t$.  This amounts to a coarse time-$\delta t$ map; the procedure is
then repeated.  The resulting scheme has already been used with
lattice-Boltzmann simulations of the Fitzhugh--Nagumo dynamics
\cite{Kevr00,KevrGearHymKevrRunTheo02} and with particle-based
simulations of the viscous Burgers equation \cite{GearLiKevr03}.

To increase the efficiency of time integration,  one can use the gap-tooth
scheme in conjuction with any method-of-lines time integration method, such as
projective integration \cite{GearKevr01}.  We then perform a number of
gap-tooth steps of size $\delta t$ to obtain an estimate of the time derivative
of the unavailable macroscopic equation.  This estimate is subsequently used to
perform a time step of size $\Delta t \gg \delta t$.  This combination has been
termed \emph{patch dynamics} \cite{KevrGearHymKevrRunTheo02}.

In this paper, we will study the patch dynamics scheme for a model diffusion
homogenization problem.  Here, the microscopic equation is a diffusion equation
with a spatially periodic diffusion coefficient with small spatial period
$\epsilon$, while the macroscopic (effective) equation describes the averaged
behaviour.  In the limit of $\epsilon$ going to zero, this effective equation
is the classical  homogenized equation.  Our goal is to approximate the
effective equation by using only the microscopic equation in a set of small
boxes.  In \cite{SamRooKevr03}, we already studied the gap-tooth scheme for
periodic reaction-diffusion homogenization problems.  We showed that the
gap-tooth scheme approximates a finite difference scheme for the homogenized
equation, when the averaged gradient is constrained at the box boundaries.
However, generally, a given microscopic code only allows us to run with a set
of predefined boundary conditions.  It is highly non-trivial to impose
macroscopically inspired boundary conditions on such microscopic codes, see
e.g.~\cite{LiYip98} for a control-based strategy.  Therefore, we circumvent
this problem here by introducing buffer regions at the boundary of each small
box, which shield the \emph{short-term} dynamics within the computational
domain of interest from boundary effects.  One then uses the microscopic code
with its \emph{built-in} boundary conditions.  In this paper, we study the
resulting \emph{gap-tooth scheme with buffers}, which was introduced in
\cite{SamKevrRoo03,SamRooKevr03}, when used inside a patch dynamics scheme, and
analyze the relation between buffer size, time step and accuracy for a model
diffusion homogenization problem.  The analysis in this context is important,
because we can clearly show the influence of the microscopic scales on the
accuracy of the solution for this model problem.  However, we emphasize that
the real advantage of the method lies in its applicability for non-PDE
microscopic simulators, e.g.~kinetic Monte Carlo or molecular dynamics.

It is worth mentioning that many numerical schemes have been devised for the
homogenization problem. Hou and Wu developed the multi-scale finite element
method that uses special basis functions to capture the correct microscopic
behaviour \cite{HouWu97,HouWu99}.  Schwab, Matache and Babuska have devised a
generalized FEM method based on a two-scale finite element space
\cite{MatBabSchwab00,SchwabMat02}.  Runborg et al.~\cite{RunTheoKevr02}
proposed a time-stepper based method that obtains the effective behaviour
through short bursts of detailed simulations appropriately averaged over many
shifted initial conditions.  The simulations were performed over the whole
domain, but the notion of effective behaviour is identical.  In their recent
work, E and Engquist and collaborators address the same problem of simulating
only the macroscopic behaviour of a multiscale model, see 
e.g.~\cite{AbdE03,EEng03}.  In what they call the heterogeneous multiscale
method, a macroscale solver is combined with an estimator for quantities that
are unknown because the macroscopic equation is not available.  This estimator
subsequently uses appropriately constrained runs of the microscopic model
\cite{EEng03}.  It should be clear that patch dynamics does exactly this: by
taking a few gap-tooth steps, we estimate the time derivative of the unknown
effective equation, and give this as input to an ODE solver, such as projective
integration.  The difference in their work is that, for conservation laws, the
macro-field time derivative is estimated from the \emph{flux} of the conserved
quantity; their generalized Godunov scheme is based on this principle.

This paper is organized as follows.  In section \ref{sec:problem}, we describe
the model homogenization problem. In section \ref{sec:dudt}, we show how to use
the gap-tooth scheme to approximate the time derivative of the unavailable
macroscopic equation. We prove a consistency result and propose a simple
heuristic to obtain a sufficient buffer size.  We also discuss to what extent
the results depend on the specific setting of our model problem.  In section
\ref{sec:patch}, we describe the full patch dynamics algorithm and give some
comments on stability.  Section \ref{sec:numeric} contains some numerical
examples which illustrate the accuracy and efficiency of the proposed method,
and we conclude in section \ref{sec:conclusions}.

\section{The homogenization problem\label{sec:problem}}

As a model problem, we consider the following parabolic partial differential
equation, 
\begin{equation}\label{eq:model_equation} 
\begin{split} 
\partial_t u_{\epsilon}(x,t)=
\partial_x\left(a\left(x/\epsilon\right)\partial_x
u_{\epsilon}(x,t) \right), \text{ in } [0,T)\times [0,1]\\
u_{\epsilon}(x,0)=u^0(x) \in L^2([0,1]),\qquad
u_{\epsilon}(0,t)=u_{\epsilon}(1,t)=0, 
\end{split} 
\end{equation} 
where $a(y)=a\left(x/\epsilon\right)$ is uniformly elliptic and periodic in $y$ and
$\epsilon$ is a small parameter.  We choose homogeneous Dirichlet boundary
conditions for simplicity.

According to classical homogenization theory \cite{BenLioPap78}, the solution
to (\ref{eq:model_equation}) can be written as an asymptotic expansion in
$\epsilon$, 
\begin{equation}
\label{eq:asymptotic_expansion}
u_{\epsilon}(x,t)=u_0(x,t)+
\sum_{i=1}^{\infty}\epsilon^i \left(u_i(x,x/\epsilon,t)\right), 
\end{equation} 
where the functions $u_i(x,y,t)\equiv u_i(x,x/\epsilon,t)$, $i=1,2,\ldots$ are
periodic in $y$.  Here, $u_0(x,t)$ is the solution of the \emph{homogenized
equation} 
\begin{equation}\label{eq:hom_eq} 
\begin{split} 
\partial_t u_0(x,t)=\partial_x\left(a^*\partial_x u_0(x,t)\right), 
\text{ in } [0,T)\times [0,1]\\ 
u_0(x,0)=u^0(x) \in L^2([0,1]), \qquad u_0(0,t)=u_0(1,t)=0,
\end{split} 
\end{equation} 
the coefficient $a^*$ is the constant effective coefficient, given by
\begin{equation} 
a^*=\int_0^1 a(y)\left(1-\frac{\d}{\d y}\chi(y)\right)\d y,
\end{equation} 
and $\chi(y)$ is the periodic solution of
\begin{equation}
\label{eq:cell_problem} 
\frac{\d}{\d y}\left(a(y)\frac{\d}{\d y}\chi(y)\right)=\frac{\d}{\d y}a(y), 
\end{equation} 
the so-called \emph{cell problem}.  The solution of (\ref{eq:cell_problem}) is
only defined up to an additive constant, so we impose the extra condition
\begin{displaymath} 
\int_0^1 \chi(y)\d y=0.  
\end{displaymath} 
From this cell problem, we can derive $u_1(x,y,t)=\partial_x u_0\chi(y)$.  We
note that in one space dimension, an explicit formula is known for $a^*$,
\cite{BenLioPap78}, 
\begin{equation} 
a^*=\left[\int_{0}^{1}\frac{1}{a(y)}\d y\right]^{-1}.  
\end{equation}

These asymptotic expansions have been rigorously justified in the classical
book \cite{BenLioPap78}, see also \cite{CioDon99}.  Under the assumptions made
on $a(x/\epsilon)$, one obtains strong convergence of $u_{\epsilon}(x,t)$ to
$u_{0}(x,t)$ as $\epsilon \to 0$ in $L^2([0,1])\times C([0,T))$. Indeed, we can
write 
\begin{equation}\label{eq:strong_conv}
\left\|u_{\epsilon}(x,t)-u_0(x,t)\right\|_{L^2([0,1])}\le C_0\epsilon,
\end{equation} 
uniformly in $t$.

It is important to note that the gradient of $u(x,t)$ is given by
\begin{equation} 
\partial_x u_{\epsilon}(x,t)=\partial_x u_0(x,t)
+\partial_y u_1(x,y,t)+O(\epsilon), 
\end{equation} 
from which it is clear that the micro-scale fluctuations have a strong effect
on the local detailed gradient.

Using the gap-tooth scheme, we will approximate the homogenized solution
$u_0(x,t)$ by a local spatial average, defined as 
\begin{displaymath}
U(x,t)=\mathcal{S}_h(u_{\epsilon})(x,t)=
\frac{1}{h}\int_{x-h/2}^{x+h/2}u_{\epsilon}(\xi,t)\d\xi.
\end{displaymath} 
It can easily be seen that that $U(x,t)$ is a good approximation to $u_0(x,t)$
in the following sense.  
\begin{lem}\label{lem:averaging_error} 
Consider $u_{\epsilon}(x,t)$ to be the solution of (\ref{eq:model_equation}),
and $u_{0}(x,t)$ to be the solution of the associated homogenized equation
(\ref{eq:hom_eq}).  Then, assuming 
\begin{equation}\label{eq:box_size}
h=O(\epsilon^p),\qquad p \in (0,1) , 
\end{equation} 
the difference between the homogenized solution $u_0(x,t)$ and the averaged
solution $U(x,t)$ is bounded by 
\begin{equation}\label{eq:averaging_error}
\left\|U(x,t)-u_0(x,t)\right\|_{L^{\infty}([0,1])}\le 
C_1 h^2+C_2\epsilon^{1-p}.  
\end{equation} 
\end{lem} 
For a proof, we refer to \cite[Lemma~3.1]{SamRooKevr03}. Note that this error
bound can be improved if we have more knowledge about the convergence of
$u_{\epsilon}$ to $u_0$ (e.g.~in $L^{\infty}([0,1])$).

\section{Estimation of the time derivative\label{sec:dudt}}

We devise a scheme for the evolution of the averaged behaviour $U(x,t)$, while
making only use of the given detailed equation (\ref{eq:model_equation}).
Moreover, we assume that a time integration code for (\ref{eq:model_equation})
has already been written and is available with a number of \emph{standard}
boundary conditions, such as no-flux or Dirichlet.  We also assume that the
order $d$ of the unavailable macroscopic equation (the highest spatial
derivative) is known.  A strategy to obtain this information is given in
\cite{LiKevrGearKevr03}.  So, we know that the macroscopic equation is of the
form 
\begin{equation}\label{eq:unknown} 
\partial_t U=F(U,\partial_x U,\ldots,\partial^d_x U,t), \end{equation} 
where $\partial_t$ denotes the time derivative and $\partial_x^k$ denotes the
$k$-th spatial derivative.

\subsection{The gap-tooth scheme with buffers\label{subsec:gap-tooth}}

Suppose we want to obtain the solution of (\ref{eq:unknown}) on the interval
$[0,1]$, using an equidistant, macroscopic mesh $\Pi(\Delta
x):=\{0=x_0<x_1=x_0+\Delta x<\ldots<x_N=1\}$.  For convenience, we define
a macroscopic comparison scheme, which is a space-time discretization for
(\ref{eq:unknown}) in the assumption that this equation is known.  We will
denote the numerical solution of this scheme by $U_i^n\approx
U(x_i,t_n)$.  Here, we choose as a comparison scheme a forward Euler/spatial
finite difference scheme, which is defined by 
\begin{equation}\label{eq:comparison}
U^{n+\delta}=S(U^n,t_n;\delta t)=
U^n+\delta t\; F(U^n,D^1(U^n),\ldots,D^d(U^n),t_n), 
\end{equation} 
where $D^k(U^n)$ denotes a suitable finite difference approximation for the
$k$-th spatial derivative.

\begin{figure} 
\begin{center}
\includegraphics[scale=0.5]{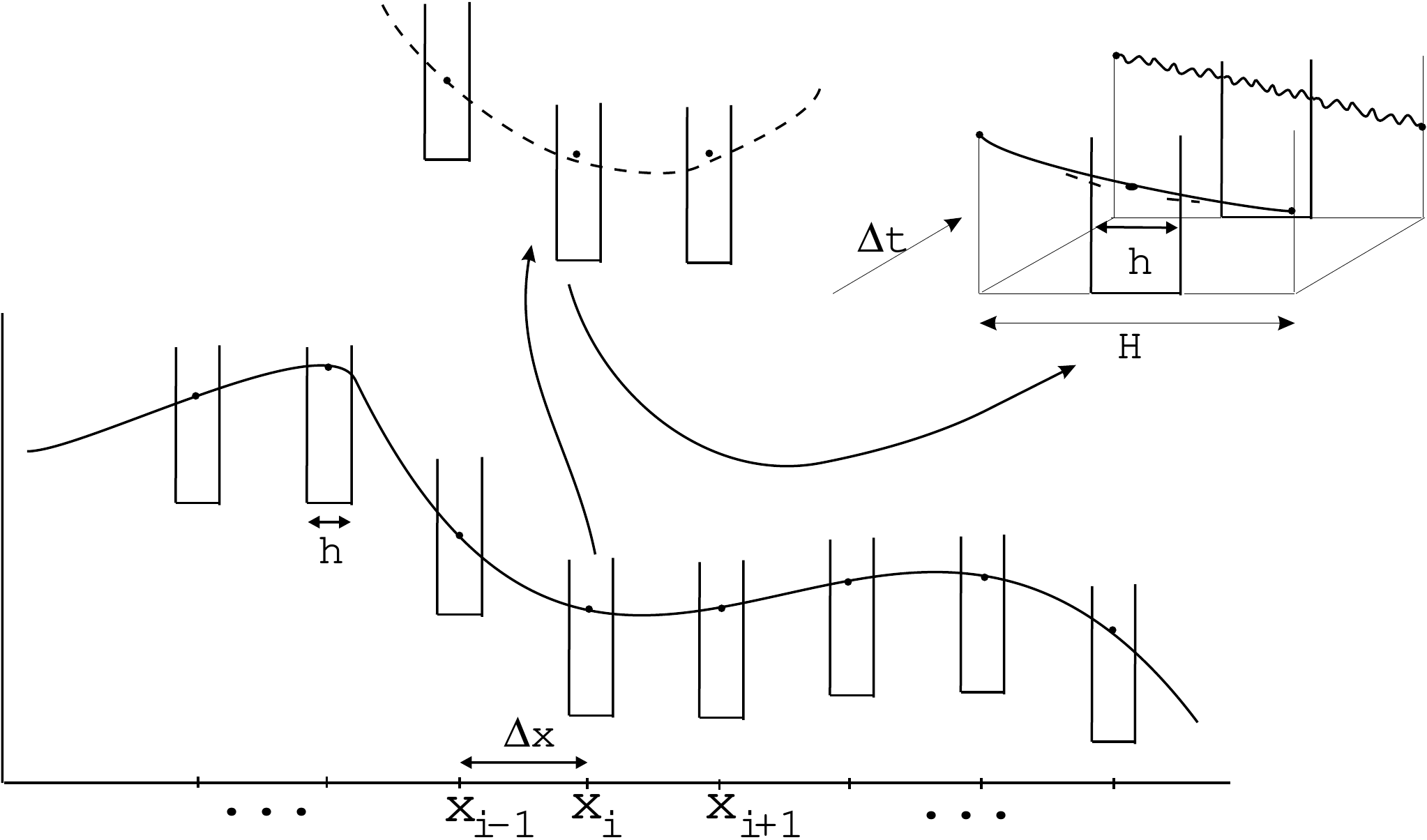}
\caption{\label{fig:gaptooth_buffers}A schematic representation of the
gap-tooth scheme with buffer boxes.  We choose a number of boxes of size $h$
around each macroscopic mesh point $x_i$ and define a local Taylor
approximation as initial condition in each box.  Simulation is performed inside
the larger (buffer) boxes of size $H$, where some boundary conditions are
imposed.} 
\end{center} 
\end{figure}

Since equation (\ref{eq:unknown}) is not known explicitly, we construct a
gap-tooth scheme to approximate the comparison scheme (\ref{eq:comparison}).
We denote the solution of the gap-tooth scheme by
$\bar{U}_i^n\approx U_i^n$.  The gap-tooth scheme is now constructed as
follows.  Consider a small interval (box, \emph{tooth}) of length $h$ around
each mesh point, as well as a larger \emph{buffer} interval of size $H>h$.
(See figure \ref{fig:gaptooth_buffers}.) We will perform a time
integration using the microscopic model (\ref{eq:model_equation}) in each box
of size $H$, and we provide this simulation with the following initial and
boundary conditions.

\paragraph{Initial condition}

We define the initial condition by constructing a local Taylor expansion, based
on the (given) box averages $\bar{U}_i^n$, $i=0,\ldots,N$, at mesh point $x_i$
and time $t_n$, 
\begin{equation}\label{eq:local_taylor}
\bar{u}^i(x,t_n)=\sum_{k=0}^{d}D^k_i(\bar{U}^n)\frac{(x-x_i)^k}{k!},\qquad 
x \in [x_i-\frac{H}{2},x_i+\frac{H}{2}], 
\end{equation} 
where $d$ is the order of the macroscopic equation.  The coefficients
$D^k_i(\bar{U}^n)$, $k>0$ are the same finite difference approximations for the
$k$-th spatial derivative that would be used in the comparison scheme
(\ref{eq:comparison}), whereas $D_i^0(\bar{U}^n)$ is chosen such that
\begin{equation}
\frac{1}{h}\int_{x_i-h/2}^{x_i+h/2}\bar{u}^i(\xi,t_n)\d\xi=\bar{U}_i^n.
\end{equation} 
For example, when $d=2$, and using standard (second-order) central differences,
we have 
\begin{equation}\label{eq:specific_initial}
D^2_i(\bar{U}^n)=
\frac{\bar{U}^n_{i+1}-2\bar{U}^n_i+\bar{U}^n_{i-1}}{\Delta x^2}, \ 
D^1_i(\bar{U}^n)=\frac{\bar{U}^n_{i+1}-\bar{U}^n_{i-1}}{\Delta x},\
D^0_i(\bar{U}^n)=\bar{U}^n_i-\frac{h^2}{12}D_i^2(\bar{U}^n).  
\end{equation}
The resulting initial condition was used in \cite{SamRooKevr03}, where it was
derived as an interpolating polynomial for the box averages.

\paragraph{Boundary conditions} 

The time integration of the microscopic model in each box should provide
information on the evolution of the \emph{global} problem at that location in
space.  It is therefore crucial that the boundary conditions are chosen such
that the solution inside each box evolves \emph{as if it were embedded in the 
larger domain}.  We already mentioned that, in many cases, it is not possible
or convenient to impose macroscopically-inspired constraints on the microscopic
model (e.g.~as boundary conditions).  However, we can introduce a larger box of
size $H>h$ around each macroscopic mesh point, but still only use (for
macro-purposes) the evolution over the smaller, inner box.  The simulation can
subsequently be performed using any of the \emph{built-in} boundary conditions
of the microscopic code.  Lifting and (short-term) evolution (using
\emph{arbitrary} available boundary conditions) are performed in the larger
box; yet the restriction is done by processing the solution (here taking its
average) over the inner, small box only.  The goal of the additional
computational domains, the \emph{buffers}, is to buffer the solution inside the
small box from the artificial disturbance caused by the (repeatedly updated)
boundary conditions.  This can be accomplished over \emph{short enough} time
intervals, provided the buffers are \emph{large enough}; analyzing the method
is tantamount to making these statements quantitative.

The idea of a buffer region was also introduced in the multiscale finite
element method of Hou (oversampling) \cite{HouWu97} to eliminate boundary layer
effects; also Hadjiconstantinou makes use of overlap regions to couple a
particle method with a continuum code \cite{Hadji99}.  If the microscopic code
allows a choice of different types of microscopic boundary conditions,
selecting the size of the buffer may also depend on this choice.

\paragraph{The algorithm}

The complete gap-tooth algorithm to proceed from $t_n$ to $t_n+\delta t$ is
given below:

\begin{enumerate} 
\item \textbf{Lifting} At time $t_n$, construct the initial
condition $\bar{u}^i(x,t_n)$, $i=0,\ldots,N$ using the box averages
$\bar{U}_i^n$, as defined in (\ref{eq:local_taylor}).  
\item \textbf{Simulation} Compute the box solution $\bar{u}^i(x,t)$, $t>t_n$, by
solving equation (\ref{eq:model_equation}) in the interval $[x_i-H/2,x_i+H/2]$
with \emph{some} boundary conditions up to time $t_{n+ \delta}=t_n+\delta t$.
The boundary conditions can be anything that the microscopic code allows.
\item \textbf{Restriction} Compute the average
$\bar{U}^{n+\delta}_i=
1/h\int_{x_i-h/2}^{x_i+h/2}\bar{u}^i(\xi,t_{n+\delta})\d\xi$
over the \emph{inner, small box only}.  
\end{enumerate}

It is clear that this procedure amounts to a map of the macroscopic variables
$\bar{U}^n$ at time $t_n$ to the macroscopic variables at time $t_{n+\delta}$,
i.e.~a ``coarse to coarse" time $\delta t$-map.  We write this map as follows,
\begin{equation}\label{eq:gaptooth_stepper}
\bar{U}^{n+\delta}=\bar{S}^d(\bar{U}^n,t_n;\delta t,H)=\bar{U}^n+\delta t\;
\bar{F}^d(\bar{U}^n,t_n;\delta t, H), 
\end{equation} 
where we introduced the time derivative estimator
\begin{equation}\label{eq:der_est} 
\bar{F}^d(\bar{U}^n,t_n;\delta t,H)=
\frac{\bar{U}^{n+\delta}-\bar{U}^n}{\delta t}.  
\end{equation} 
The superscript $d$ denotes the highest spatial derivative that has been
prescribed by the initialization scheme (\ref{eq:local_taylor}).  The accuracy
of this estimate depends on the buffer size $H$, the box size $h$ and the time
step $\delta t$.

\subsection{Consistency}\label{subsec:consistency}

To analyze convergence, we solve the detailed problem approximately in each
box.  Because $h\gg\epsilon$, we can resort to the homogenized solution, and
bound the error using equation (\ref{eq:strong_conv}).  It is important to note
that we use the homogenized equation for analysis purposes only.  The algorithm
uses box averages of solutions of the detailed problem
(\ref{eq:model_equation}), so it does not exploit more knowledge of the
homogenized equation than the order $d$.  We choose to study convergence in the
case of Dirichlet boundary conditions, but we will show numerically that the
results do not depend crucially on the type of boundary conditions.

We first relate the gap-tooth time-stepper as constructed in section
\ref{subsec:gap-tooth} to a gap-tooth time-stepper in which the microscopic
equation has been replaced by the homogenized equation.
\begin{lem}\label{lem:equivalent_box_problem} 
Consider the model equation,
\begin{equation}\label{eq:model_eq_repeated}
\partial_t u_{\epsilon}(x,t)=\partial_x\left(a\left(x/\epsilon\right)\partial_x
u_{\epsilon}(x,t) \right), 
\end{equation} 
where $a(y)=a\left(x/\epsilon\right)$ is periodic in $y$ and $\epsilon\ll1$,
with initial condition $u_{\epsilon}(x,0)=u^0(x)$ and Dirichlet boundary
conditions 
\begin{equation}\label{eq:bc_r} 
u_{\epsilon}(-H/2,t)=u^0(-H/2), \qquad 
u_{\epsilon}(H/2,t) =u^0(H/2).
\end{equation} 
For $\epsilon\to 0$, this problem converges to the homogenized problem
\begin{equation}\label{eq:h_r} 
\partial_t u_0(x,t)=\partial_x\left(a^*\partial_x u_0(x,t)\right) 
\end{equation} 
with initial condition $u_0(x,0)=u^0(x)$ and Dirichlet boundary conditions
\begin{equation}\label{eq:h_bc_r} 
u_{\epsilon}(-H/2,t)=u^0(-H/2), \qquad 
u_{\epsilon}(H/2,t) =u^0(H/2).
\end{equation} 
and the solution of (\ref{eq:model_eq_repeated})-(\ref{eq:bc_r}) converges
pointwise to the solution of (\ref{eq:h_r})-(\ref{eq:h_bc_r}), with the
following error estimate 
\begin{equation}
\left\|u_{\epsilon}(x,t)-u_0(x,t)\right\|_{L^2([-H/2,H/2])}\le C_3\epsilon.
\end{equation} 
\end{lem} 
This is a standard result, whose proof can be found in
e.g.~\cite{All92,CioDon99}.

We now define two gap-tooth time-steppers.  Let
\begin{equation}\label{eq:gaptooth_detailed}
\bar{U}^{n+\delta}=\bar{S}^2(\bar{U}^n,t_n;\delta t, H)=\bar{U}^n+\delta t \;
\bar{F}^2(\bar{U}^n,t_n;\delta t, H) 
\end{equation} 
be a gap-tooth time-stepper that uses the detailed, homogenization problem
(\ref{eq:model_eq_repeated})-(\ref{eq:bc_r}) inside each box, and
\begin{equation}\label{eq:gaptooth_homogenized}
\hat{U}^{n+\delta}=\hat{S}^2(\hat{U}^n,t_n;\delta t, H)=\hat{U}^n+\delta t \;
\hat{F}^2(\hat{U}^n,t_n;\delta t, H) 
\end{equation} 
be a gap-tooth time-stepper where the homogenization problem for each box has
been replaced by the homogenized equation (\ref{eq:h_r})-(\ref{eq:h_bc_r}).
The box initialization is done using a quadratic polynomial as defined in
(\ref{eq:specific_initial}).

We can apply \cite[Lemma~4.2]{SamRooKevr03} to bound the difference between
$\bar{F}^2(\bar{U},t_n;\delta t,H)$ and $\hat{F}^2(\hat{U},t_n;\delta t,H)$.
\begin{lem} 
Consider $\bar{U}^{n+\delta}=\bar{S}^2(\bar{U}^{n},t_n;\delta t,H)$ and
$\hat{U}^{n+\delta}=\hat{S}^2(\hat{U}^n,t_n;\delta t,H)$ as defined in
(\ref{eq:gaptooth_detailed}) and (\ref{eq:gaptooth_homogenized}), respectively.
Assuming $\bar{U}^n=\hat{U}^n$, $h=O(\epsilon^{p})$, $p\in (0,1)$, $\epsilon
\to 0$, we have 
\begin{displaymath} 
\left\|\bar{U}_i^{n+\delta}-\hat{U}_i^{n+\delta}\right\|\le
C_4\epsilon^{1-p/2}, \end{displaymath} 
and therefore 
\begin{displaymath} 
\left\|\bar{F}^2(\bar{U}^n,t_n;\delta t,H)-
\hat{F}^2(\hat{U}^n,t_n;\delta t,H)\right\| 
\le C_4\frac{\epsilon^{1-p/2}}{\delta t}.  
\end{displaymath}
\end{lem} 
Again, note that the error estimate can be made sharper if additional knowledge
of the convergence of $u_{\epsilon}$ to $u_0$ is available.

It can easily be checked that the averaged solution $U(x,t)$ also satisfies the
diffusion equation (\ref{eq:h_r}).  Therefore, we define the comparison scheme
(\ref{eq:comparison}) for the model problem as 
\begin{eqnarray}
U^{n+\delta}&=&S(U^n,t_n;\delta t) \nonumber\\ 
&=&U^n+\delta t \; F(U^n,D^1(U^n),D^2(U^n),t_n)\nonumber\\ 
&=&U^n+\delta t \left[ a^*\;D^2(U^n)\right].\label{eq:comparison_hom} 
\end{eqnarray}

The following theorem compares the gap-tooth time derivative estimator
$\hat{F}^2(\hat{U}^n,t_n;\delta t,H)$ with the finite difference time
derivative used in (\ref{eq:comparison_hom}).
\begin{thm}\label{thm:diff_wrt_fd} 
Consider the gap-tooth time-stepper for the homogenized equation, as defined by
(\ref{eq:gaptooth_homogenized}), and the corresponding comparison scheme
(\ref{eq:comparison_hom}).  Assuming $U^n=\hat{U}^n$, and defining the error
\begin{displaymath} 
E(\delta t,H)=
\left\|\hat{F}^2(\hat{U}^n,t_n;\delta t,H) -a^*\; D^2(U^n)\right\|, 
\end{displaymath} 
we have the following result for $\delta t/H^2 \to 0$, $h\ll H$, 
\begin{equation}\label{eq:error_bound} 
E(\delta t,H)\le\left(C_1+C_2\frac{h^2}{\delta t}\right)
\left(1-\exp(-a^*\pi^2\frac{\delta t}{H^2})\right) 
\end{equation}
\end{thm} 
\begin{proof} 
First, we solve the equation (\ref{eq:h_r})-(\ref{eq:h_bc_r}) analytically
inside each box, with initial condition given by (\ref{eq:local_taylor}).
Using the technique of separation of variables, we obtain \begin{displaymath}
\begin{split}
\hat{u}^i(x,t)=&\hat{U}^n_i-\frac{h^2}{12}D_i^2(\hat{U}^n)+
D_i^2(\hat{U}^n)\frac{H^2}{8}+D_i^1(\hat{U}^n)(x-x_i)\\
+&\sum_{m=1}^{\infty}a_m^i\exp\left(-a^*\frac{m^2\pi^2}{H^2}(t-t_n)\right)
\sin\left(\frac{m\pi}{H}(x-x_i-\frac{H}{2})\right), 
\end{split}
\end{displaymath} 
where 
\begin{displaymath}
a_m^i=\frac{2}{H}\int_{x_i-H/2}^{x_i+H/2}
\frac{1}{2}D_i^2(\hat{U}^n)\left((x-x_i)^2-\frac{H^2}{4}\right)
\sin\left(\frac{m\pi}{H}(x-x_i-\frac{H}{2})\right)\d x.  
\end{displaymath} 
This can be simplified to 
\begin{displaymath} 
a_m^i=-\frac{2H^2 D_i^2(\hat{U}^n)\left((-1)^{m}-1\right)}{m^3 \pi^3}, 
\end{displaymath} 
which yields the following solution, \begin{multline}
\hat{u}^i(x,t)=\hat{U}_i^n-\frac{h^2}{12}D_i^2(\hat{U}^n)
+D^2_i(\hat{U}^n)\frac{H^2}{8}+D_i^1(\hat{U}^n) (x-x_i)\\ 
+\sum_{m=1}^{\infty}\frac{4H^2 D_i^2(\hat{U}^n)}{(2m-1)^3\pi^3}
\exp\left(-a^*\frac{(2m-1)^2\pi^2}{H^2}(t-t_n)\right)
\sin\left(\frac{(2m-1)\pi}{H}(x-x_i-\frac{H}{2})\right).
\end{multline} 
When taking the average over a box of size $h$, we obtain,
\begin{multline}
\frac{1}{h}\int_{x_i-h/2}^{x_i+h/2}\hat{u}^i(x,t)\d x=
\hat{U}_i^n-\frac{h^2}{12}D_i^2(\hat{U}^n)+D^2_i(\hat{U}^n)\frac{H^2}{8}\\
+\sum_{m=1}^{\infty}\frac{4H^2\alpha_m}{(2m-1)^3\pi^3}D_i^2(\hat{U}^n)
\exp\left(-a^*\frac{(2m-1)^2\pi^2}{H^2}(t-t_n)\right),
\end{multline} 
with $\alpha_m$ determined by 
\begin{eqnarray*}
\alpha_m&=&\frac{1}{h}\int_{x_i-h/2}^{x_i+h/2}
\sin\left(\frac{(2m-1)\pi}{H}(x-x_i-\frac{H}{2})\right)\d x\\
&=&\frac{H}{(2m-1)h\pi}\left(\cos\left(\frac{(2m-1)\pi}{2H}(H+h)\right)
-\cos\left(\frac{(2m-1)\pi}{2H}(H-h)\right)\right)\\
&=&(-1)^m\frac{2H}{(2m-1)h\pi}\sin\left(\frac{(2m-1)\pi}{2H}h\right).
\end{eqnarray*} 
The coefficients $\alpha_m$ tend to $1$ in absolute value as $h\to 0$.  To
obtain the time derivative estimate $\hat{F}(\hat{U}^n,t_n;\delta t,H)$, we
proceed as follows: 
\begin{eqnarray*} 
\hat{F}^2_i(\hat{U}^n,t_n;\delta t,H)&=&\frac{1}{\delta t\; h}
\int_{x_i-h/2}^{x_i+h/2}\hat{u}^i(x,t_n+\delta t)-\hat{u}^i(x,t_n)\;\d x\\ 
&=&\frac{1}{\delta t}\sum_{m=1}^{\infty}\frac{4H^2\alpha_m}{(2m-1)^3\pi^3}
D_i^2(\hat{U}^n)\left(
\exp\left(-a^*\frac{(2m-1)^2\pi^2}{H^2}\delta t\right)-1\right)\\ 
&=&4 D_i^2(\hat{U}^n) \frac{1}{\xi}\left(\sum_{m=1}^{\infty}
\frac{\alpha_m\left(
\exp\left(-a^*(2m-1)^2\pi^2\xi\right)-1\right)}{(2m-1)^3\pi^3}\right)
\end{eqnarray*} 
where we introduced $\xi=\delta t/H^2$.  It can easily be checked (e.g.~using
Maple) that 
\begin{displaymath} 
\lim_{\xi\to 0}\hat{F}^2_i(\hat{U}^n,t_n;\delta t,H)=a^* D_i^2(\hat{U}^n), 
\end{displaymath} 
which already shows that the gap-tooth scheme is consistent in this limit.
Obtaining an error bound in terms of $\xi$ is somewhat more involved.  We split
$\hat{F}_i^2(\hat{U}^n,t_n;\delta t,H)$ as follows, 
\begin{displaymath}
\hat{F}(\hat{U}^n,t_n;\delta t,H)=\hat{F}_1+\hat{F}_2, 
\end{displaymath} 
with $\hat{F}_1$ and $\hat{F}_2$ defined as 
\begin{eqnarray*} 
\hat{F}_1&=&4 D_i^2(\hat{U}^n) \frac{1}{\xi}\left(\sum_{m=1}^{\infty}
\frac{(-1)^m\left(
\exp\left(-a^*(2m-1)^2\pi^2\xi\right)-1\right)}{(2m-1)^3\pi^3}\right)\\
\hat{F}_2&=&4 D_i^2(\hat{U}^n) \frac{1}{\xi}\left(\sum_{m=1}^{\infty}
\frac{(\alpha_m-(-1)^m)\left(
\exp\left(-a^*(2m-1)^2\pi^2\xi\right)-1\right)}{(2m-1)^3\pi^3}\right).
\end{eqnarray*} 
We now show that $\hat{F_1}$ approaches the correct estimate
exponentially.  Some algebraic manipulation results in 
\begin{eqnarray*}
\hat{F}_1-a^* D_i^2(\hat{U}^n)&=&4 D_i^2(\hat{U}^n)
\frac{1}{\xi}\left(\sum_{m=1}^{\infty}\frac{(-1)^m\left(
\exp\left(-a^*(2m-1)^2\pi^2\xi\right)-1\right)}{(2m-1)^3\pi^3}
\right)-a^*D_i^2(\hat{U}^n)\\
&=&4D_i^2(\hat{U}^n)\sum_{m=1}^{\infty}
(-1)^{m+1}\left(\frac{1-a^*(2m-1)^2\pi^2\xi-
\exp\left(-a^*(2m-1)^2\pi^2\xi\right)}{(2m-1)^3\pi^3\xi}\right)
\end{eqnarray*} 
Therefore, we have 
\begin{multline*} 
\left\|\hat{F}_1-a^* D_i^2(\hat{U}^n)\right\|=\\
\left\|4D_i^2(\hat{U}^n)\sum_{m=1}^{\infty}(-1)^{m+1}a^*
\left(\frac{1-a^*(2m-1)^2\pi^2\xi-\exp\left(-a^*(2m-1)^2\pi^2\xi\right)}
{a^*(2m-1)^3\pi^3\xi}\right)\right\|
\end{multline*} 
\begin{eqnarray} \phantom{ \left\|\hat{F}_1-a^*D_i^2(\hat{U}^n)\right\|}
&\le&C\left(\frac{1-a^*\pi^2\xi-
\exp\left(-a^*\pi^2\xi\right)}{a^*\pi^2\xi}\right)\nonumber\\
&\le&C\left(1-\exp\left(-a^*\pi^2\xi\right)\right)\label{eq:proof_first}
\end{eqnarray} 
It remains to show the asymptotic behaviour of $\hat{F}_2$.
\begin{multline*} 
\left\|\hat{F}_2\right\| = \\
\left\|4D_i^2(\hat{U}^n)\sum_{m=1}^{\infty}(-1)^m
\frac{\left(\sin\left(\frac{(2m-1)\pi h}{2H}\right)\frac{2H}{(2m-1)\pi
h}-1\right)}{(2m-1)^3\pi^3}
\left(\exp\left(-a^*\left(2m-1\right)^2\pi^2\xi\right)-1\right)\right\|
\end{multline*} 
\begin{eqnarray} 
&\le&C\left(\frac{\sin\left(\frac{\pi h}{2H}\right)
\frac{2H}{\pi h}-1}{\xi\pi^3}\right)
\left(1-\exp\left(-a^*\pi^2\xi\right)\right)\nonumber\\ 
&\le& C \frac{h^2}{H^2}\frac{1-\exp(-a^*\pi^2\xi)}{\xi\pi^3}\nonumber\\ 
&\le& C \frac{h^2}{\delta t}\left(1-\exp(-a^*\pi^2\xi)\right)
\label{eq:proof_second}
\end{eqnarray} 
The combination of (\ref{eq:proof_first}) and (\ref{eq:proof_second}) proves
the theorem.  
\end{proof} 
The error bound (\ref{eq:error_bound}) clearly shows
an exponential decay of the error as a function of $\delta t/H^2$ when the
microscopic problem is replaced by the effective equation.  The restriction
(here performed by taking the box average) also affects the accuracy of the
estimate.  Ideally, one would just use the effective function value at $x=x_i$
inside each box (this corresponds to $h=0$), but when microscopic scales are
present, this value is generally impossible to obtain.

We illustrate this result numerically.  
\begin{ex} Consider the model problem (\ref{eq:h_r}) with $a^*=0.45825686$ as a
microscopic problem on the domain $[0,1]$ with homogeneous Dirichlet boundary
conditions and initial condition $u(x,0)=1-4(x-1/2)^2$.  To solve this
microscopic problem, we use a second order finite difference discretization
with mesh width $\delta x=2\cdot 10^{-7}$ and \verb#lsode# as time-stepper.
The concrete gap-tooth scheme for this example is defined by the initialization
(\ref{eq:specific_initial}).  We compare a gap-tooth step with $h=2\cdot
10^{-3}$ and $\Delta x=1\cdot 10^{-1}$ with the reference estimator $a^*
D^2(\hat{U}^n)$.  Figure \ref{fig:error_estimate_ex3.4} shows the error with
respect to the finite difference time derivative as a function of $H$ (left)
and $\delta t$ (right).  It is clear the convergence is in agreement with
theorem \ref{thm:diff_wrt_fd}.  The stagnation for large buffer sizes is due to
the finite accuracy of the microscopic solver.  
\begin{figure}
\subfigure{\includegraphics[scale=0.7]{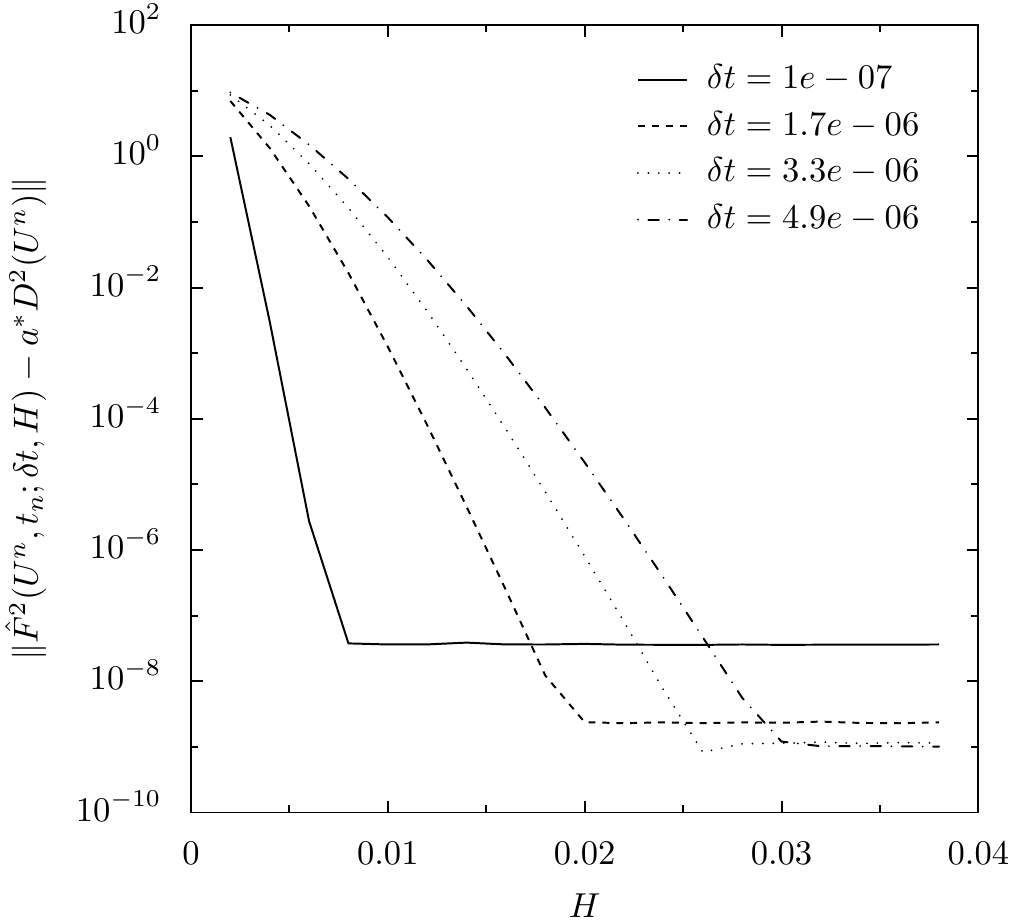}}
\subfigure{\includegraphics[scale=0.7]{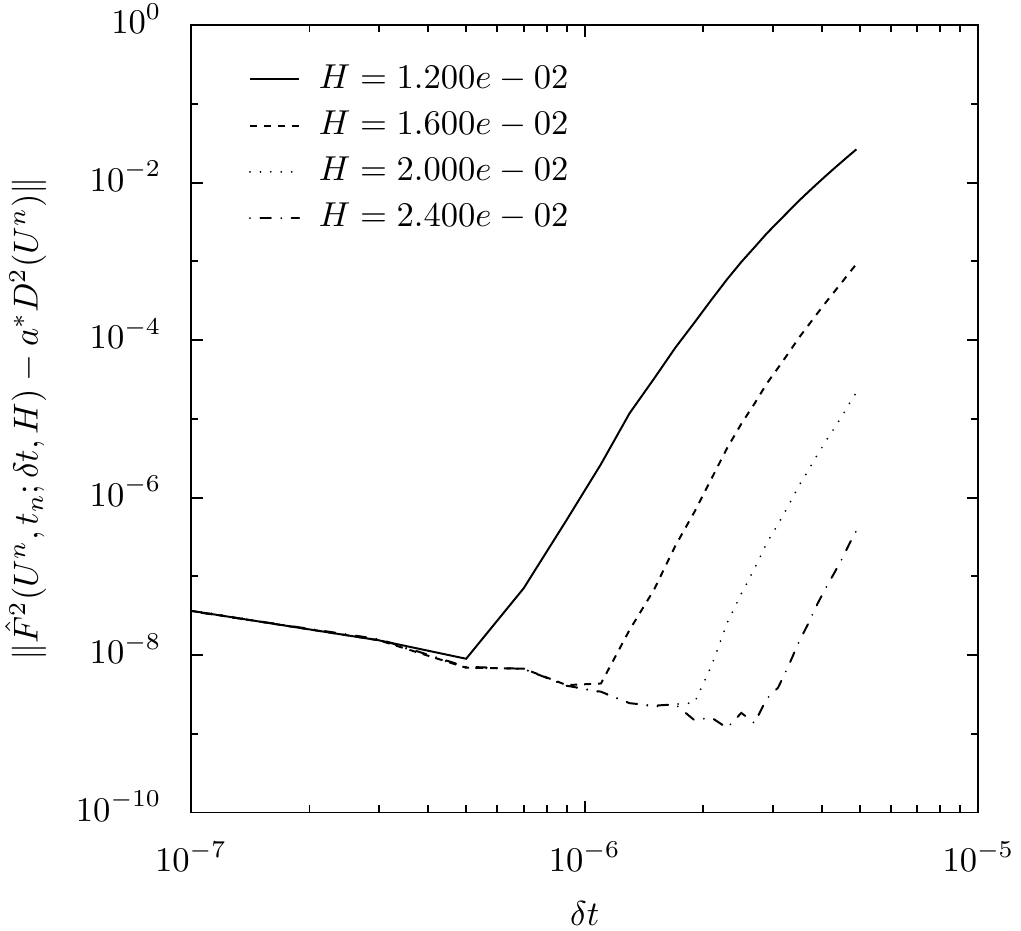}}
\caption{Error of the gap-tooth estimator $\hat{F}^2(U^n,t_n;\delta t,H)$
(which uses the homogenized problem (\ref{eq:h_r})-(\ref{eq:h_bc_r}) inside
each box) with respect to the finite difference time derivative $a^*D^2(U^n)$
on the same mesh. Left: Error with respect to $H$ for fixed $\delta t$.  Right:
Error with respect to $\delta t$ for fixed $H$.
\label{fig:error_estimate_ex3.4}} \end{figure} 
\end{ex}

We are now ready to state the general consistency result.
\begin{thm}\label{thm:consistency} 
Let $\bar{U}^{n+\delta}=\bar{S}^2(\bar{U}^n,t_n;\delta_t,H)$ be a gap-tooth
time-stepper for the homogenization problem
(\ref{eq:model_eq_repeated})-(\ref{eq:bc_r}), as defined in
(\ref{eq:gaptooth_detailed}), and $U^{n+\delta}=S(U^n,t_n;\delta t)$ a
comparison finite difference scheme as defined in (\ref{eq:comparison_hom}).
Then, assuming $U^n=\bar{U}^n$, we have, 
\begin{equation}\label{eq:consistency}
\left\|\bar{F}^2(\bar{U}^n,t_n;\delta t,H)-a^*D^2(U^n)\right\| 
\le C_4 \underbrace{
\frac{\epsilon^{1-p/2}}{\delta t}}_{\textrm{microscales}}
+C_5\underbrace{\left(1+\frac{h^2}{\delta t}\right)}_{\textrm{averaging}}
\underbrace{\left(1-\exp(-a^*\pi^2\frac{\delta t}{H^2})\right)}_{
\textrm{boundary conditions}} 
\end{equation}
\end{thm} 
\begin{proof} 
This simply follows by combining theorem \ref{thm:diff_wrt_fd} with lemma
\ref{lem:equivalent_box_problem}.  
\end{proof}

Formula (\ref{eq:consistency}) shows the main consistency properties of the
gap-tooth estimator.  The error decays exponentially as a function of buffer
size, but the optimal accuracy of the estimator is limited by the presence of
the microscopic scales.  Therefore, we need to make a trade-off to determine an
optimal choice for $H$ and $\delta t$.  The smaller $\delta t$, the smaller $H$
can be to reach optimal accuracy (and thus the smaller the compational cost),
but smaller $\delta t$ implies a larger optimal error.  This is illustrated in
the following numerical example.  
\begin{ex}\label{ex:3.6} 
Consider the model problem (\ref{eq:model_eq_repeated}) with
\begin{equation}\label{eq:diff_coeff}
a(x/\epsilon)=1.1+\sin(2\pi x/\epsilon) ,\qquad \epsilon=1\cdot 10^{-5}
\end{equation}
as a microscopic problem on the domain $[0,1]$ with homogeneous Dirichlet
boundary conditions and initial condition $u(x,0)=1-4(x-1/2)^2$.  This
diffusion coefficient has also been used as a model example in
\cite{AbdE03,SamRooKevr03}.  To solve this microscopic problem, we use a second
order finite difference discretization with mesh width $\delta x=1\cdot
10^{-7}$ and \verb#lsode# as time-stepper.  The concrete gap-tooth scheme for
this example is defined by the initialization (\ref{eq:specific_initial}).  We
compare a gap-tooth step with $h=2\cdot 10^{-3}$ and $\Delta x=1\cdot 10^{-1}$
with the reference estimator $a^* D^2(\hat{U}^n)$, in which the effective
diffusion coefficient is known to be $a^*=0.45825686$.  Figure
\ref{fig:error_estimate_ex3.6} shows the error with respect to the finite
difference time derivative as a function of $H$ (left) and $\delta t$ (right).
It is clear that the convergence is in agreement with theorem
\ref{thm:consistency}.  We see that smaller values of $\delta t$ result in
larger values for the optimal error, but the convergence towards this optimal
error is faster.  
\begin{figure}
\subfigure{
\includegraphics[scale=0.7]{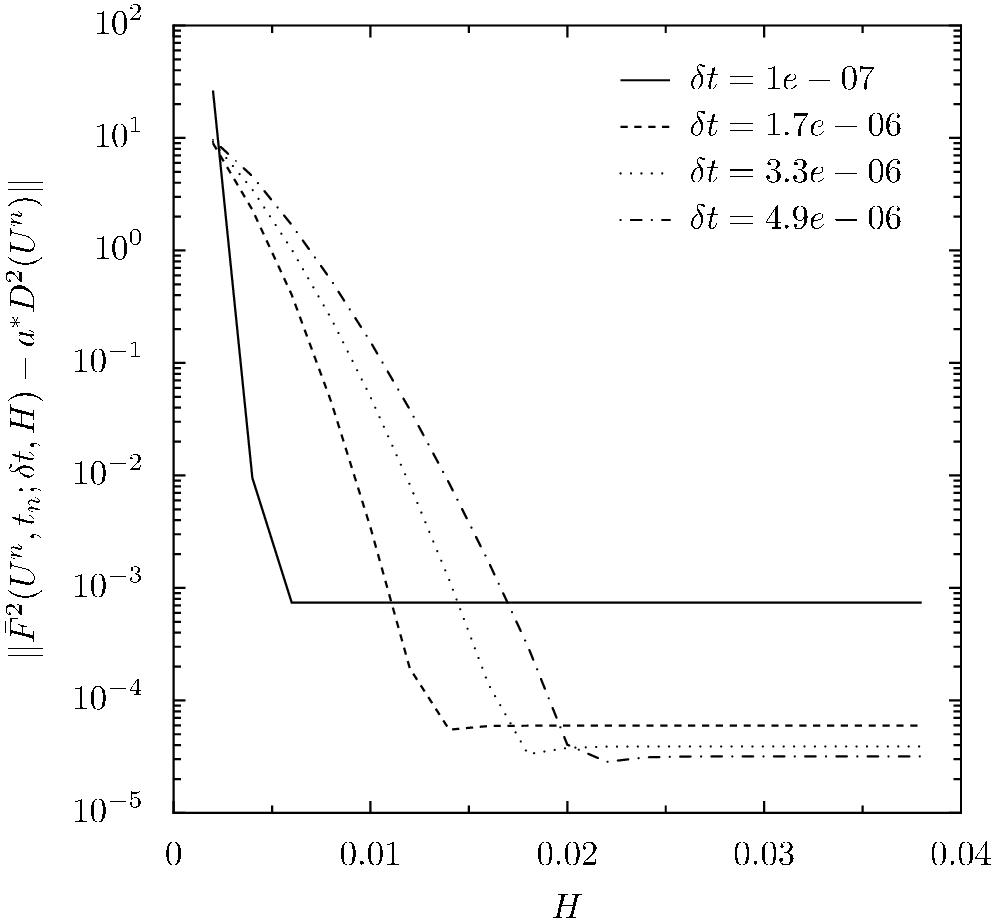}}
\subfigure{
\includegraphics[scale=0.7]{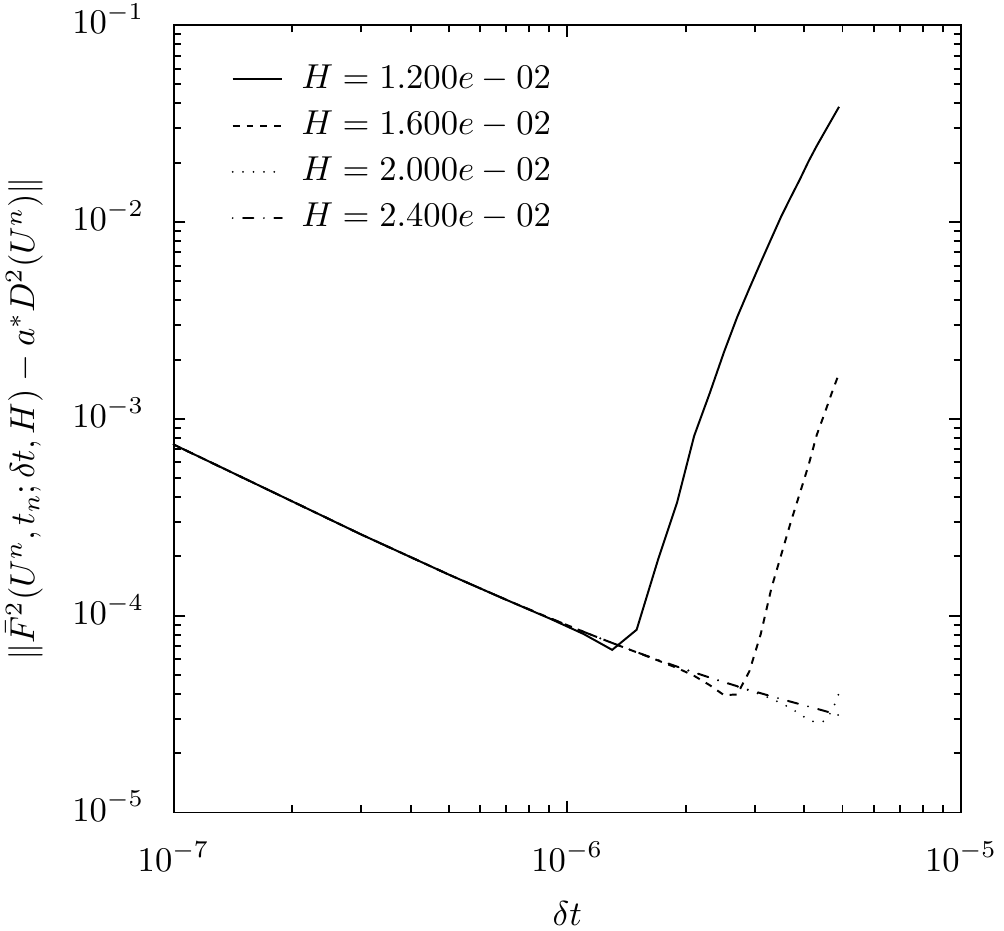}}
\caption{Error of the gap-tooth estimator $\bar{F}(U^n,t_n;\delta t,H)$ (which
uses the detailed, homogenization problem
(\ref{eq:model_eq_repeated})--(\ref{eq:bc_r}) inside each box) with respect to
the finite difference time derivative $a^*D^2(U^n)$ on the same mesh.  Left:
Error with respect to $H$ for fixed $\delta t$.  Right: Error with respect to
$\delta t$ with fixed $H$.  \label{fig:error_estimate_ex3.6}} \end{figure}
\end{ex}

\subsection{Choosing the method parameters}

When performing time integration using patch dynamics, one must determine a
macroscopic mesh width $\Delta x$, an inner box size $h$, a buffer box size $H$
and a time step $\delta t$.  These method parameters need to be chosen
adequately to ensure an accurate result.  Since the gap-tooth estimator
approximates the time derivative that would be obtained through a
method-of-lines discretization of the macroscopic equation, the macroscopic
mesh width $\Delta x$ can be determined by macroscopic properties of the
solution only, enabling reuse of existing remeshing techniques for PDEs.  The
box width $h$ has to be sufficiently large to capture all small scale effects,
but small enough to ensure a good spatial resolution. Here, we just choose
$h\gg\epsilon$.  In our simplified setting, where the microscopic model is also
a partial differential equation, we are free to choose $\delta t$, which allows
us to illustrate the convergence properties of the method.  However, in
practical problems, the choice of $\delta t$ will be problem-dependent, since
it will need to be chosen large enough to deduce reliable information on the
macroscopic time derivative.
  
\begin{figure} 
\begin{center}
\subfigure{\includegraphics[scale=0.7]{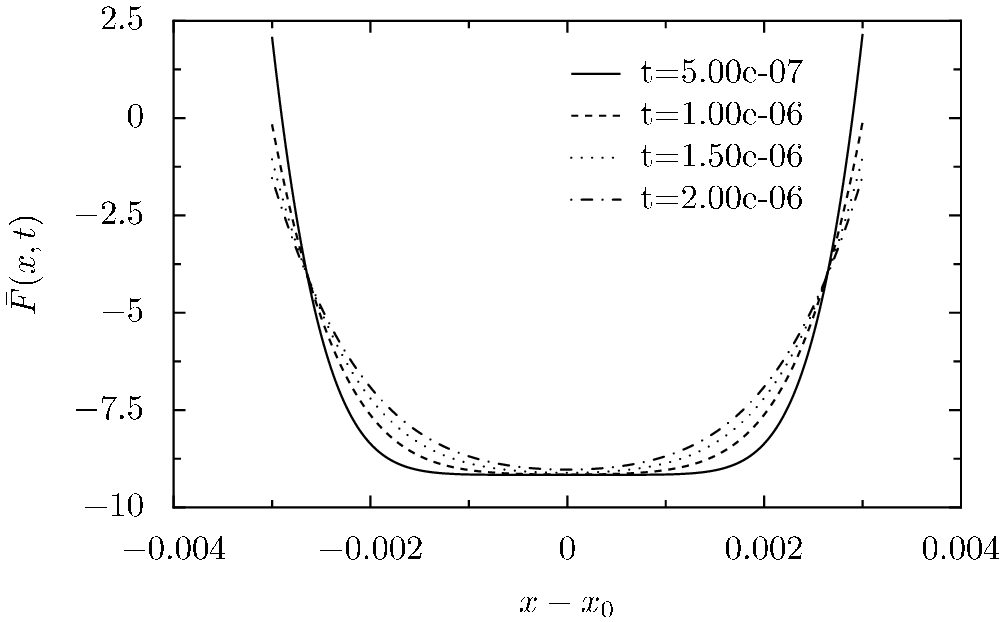}}
\subfigure{\includegraphics[scale=0.7]{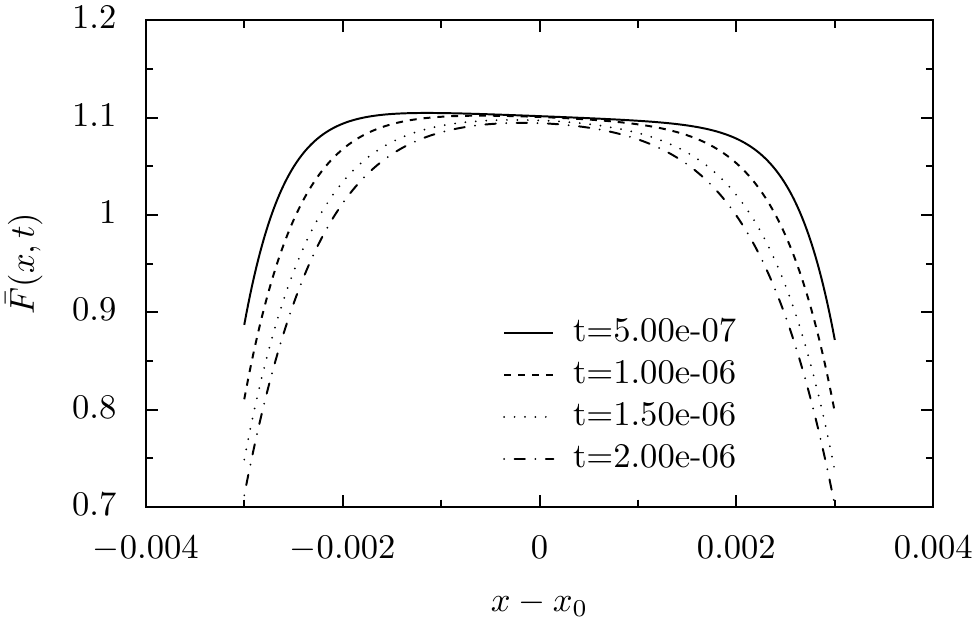}}
\caption{\label{fig:f_avg_ifv_t}The function $\bar{F}(x,t)$ as defined in
equation (\ref{eq:est_ifv_x}) for a number of values of time, using a buffer
size $H=8\cdot 10^{-3}$ and $h=2\cdot 10^{-3}$.  Left: the model diffusion
problem (\ref{eq:model_eq_repeated}-\ref{eq:bc_r}).  Right: the
reaction-diffusion equation \ref{eq:rd_hom}.  The estimate clearly gets
affected by the boundary conditions as time advances.} 
\end{center} 
\end{figure}

\begin{figure}
\begin{center}
\includegraphics{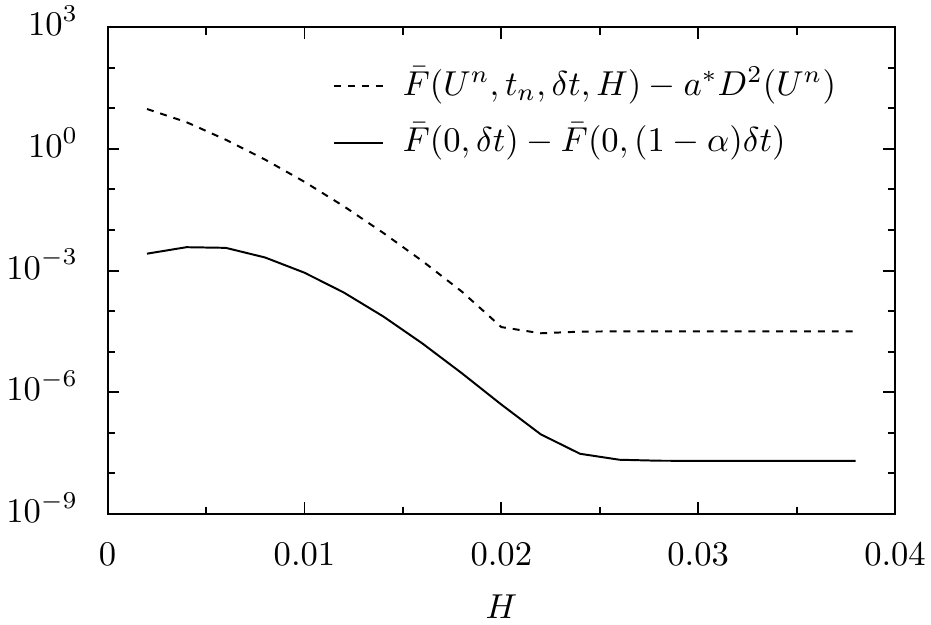}
\caption{\label{fig:test_heuristic} Error of the gap-tooth estimator (dashed)
and heuristic error estimate (solid) as a function of buffer size for the model
equation (\ref{eq:model_eq_repeated}) with diffusion coefficient
(\ref{eq:diff_coeff}) for $\delta t = 5\cdot 10^{-6}$ and $\alpha=0.04$.} 
\end{center} 
\end{figure}

Therefore, we focus on determining the buffer width $H$, assuming that all other
parameters have already been fixed.  From theorem \ref{thm:consistency}, it
follows that the desired value of $H$ depends on the effective diffusion
coefficient $a^*$, which is unknown.  We thus need to resort to a heuristic.
Consider the model problem (\ref{eq:model_eq_repeated})-(\ref{eq:bc_r})
inside one box, centered around $x_0=5\cdot 10^{-1}$, with $H=8\cdot 10^{-3}$,
with initial condition $u^0(x)=1-4(x-1/2)^2$.  The diffusion coefficient is
given by (\ref{eq:diff_coeff}), see example \ref{ex:3.6}. Denote the solution of
this problem by $\bar{u}(x,t)$, and define 
\begin{eqnarray}
\bar{F}(x,t)&=&
\mathcal{S}_h\left(\bar{u}(x,t)-\bar{u}(x,0)\right)\nonumber\\*
&=&\frac{1}{t}\int_{\xi-x_0-h/2}^{\xi-x_0+h/2}
\frac{\bar{u}(\xi,t)-\bar{u}(\xi,0)}{h}\d\xi,
\label{eq:est_ifv_x} 
\end{eqnarray} 
with $h=2\cdot 10^{-3}$ and $x\in [(-H+h)/2,(H-h)/2]$.  Figure
\ref{fig:f_avg_ifv_t} (left) shows $\bar{F}(x,t)$ for a number of values of
$t$.  We clearly see how the error in the estimator propagates inwards from the
boundaries.  The same function is plotted on the right, only now the
microscopic model is the reaction-diffusion equation 
\begin{equation}\label{eq:rd_hom}
\begin{split}
\partial_t u_{\epsilon}(x,t)=
\partial_x\left(a(x/\epsilon)\partial_x u_{\epsilon}(x,t)\right) 
+u_{\epsilon}(x,t)\left(1-\frac{u_{\epsilon}(x,t)}{1.2+\sin(2\pi x)}\right)\\ 
u_{\epsilon}(-H/2,t)=u^0(-H/2),\qquad u_{\epsilon}(H/2,t)=u^0(H/2),
\end{split}
\end{equation}
again with $a(x/\epsilon)$ defined as in (\ref{eq:diff_coeff}). In the presence
of reaction terms, $\bar{F}(x,t)$ is no longer constant in the internal region.
Based on these observations, we propose the following test for the
quality of the buffer size,
\begin{equation}\label{eq:heuristic}
\|\bar{F}(0,\delta t)-\bar{F}(0,(1-\alpha)\delta t)\|
<\verb#Tr#, \qquad 0 < \alpha \ll 1.
\end{equation}
Figure \ref{fig:test_heuristic} shows this heuristic, together with the error,
as a function of $H$ for $\delta t = 5\cdot 10^{-6}$ and $\alpha=0.04$.
It is clear that the computed quantity in (\ref{eq:heuristic}) is proportional
to the error for sufficiently large $H$.  However, this heuristic is far from
perfect, since the simulations inside each box can converge to a steady state
due to the Dirichlet boundary conditions.  If this steady state is reached in a
time interval smaller than $\delta t$, equation (\ref{eq:heuristic}) will
underestimate the error, resulting in an insufficient buffer size $H$ getting
accepted.  However, as soon as the problem-dependent parameters $\alpha$ and
\verb#Tr# have been determined, this heuristic can be used during the
simulation to check whether the currently used buffer size is still sufficient.  

\subsection{Discussion}

\begin{figure} 
\subfigure{
\includegraphics[scale=0.7]{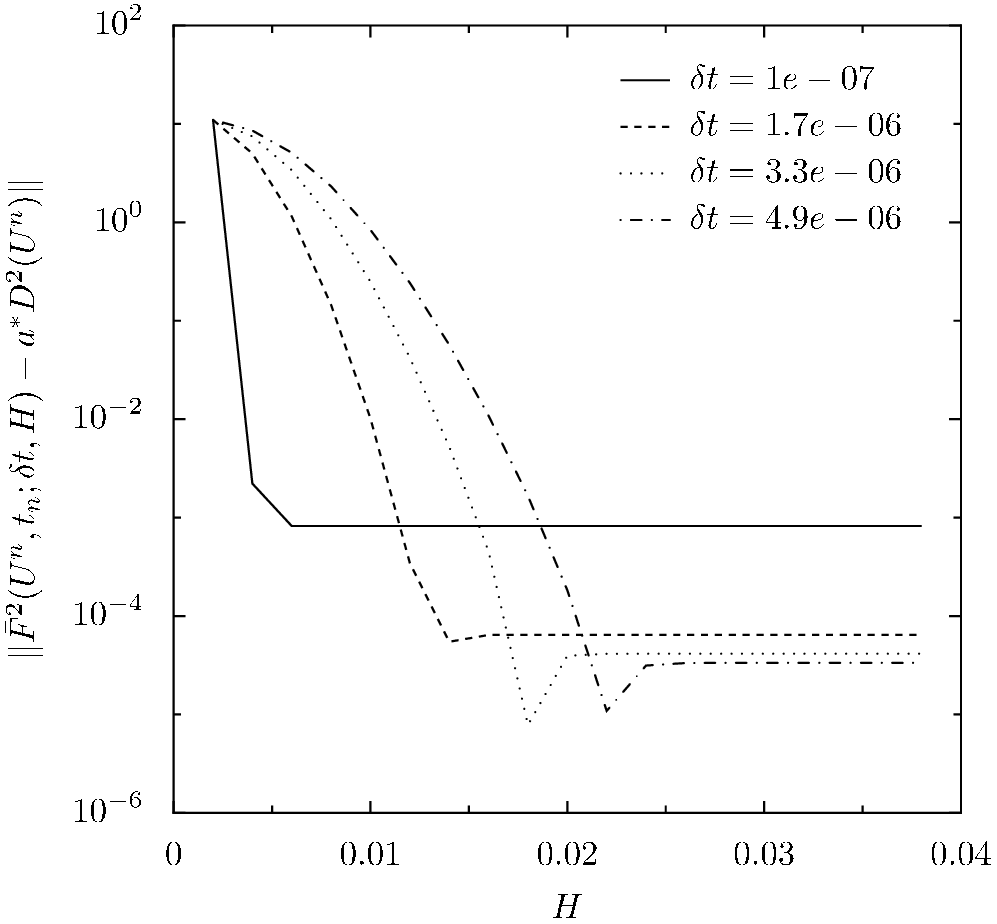}}
\subfigure{
\includegraphics[scale=0.7]{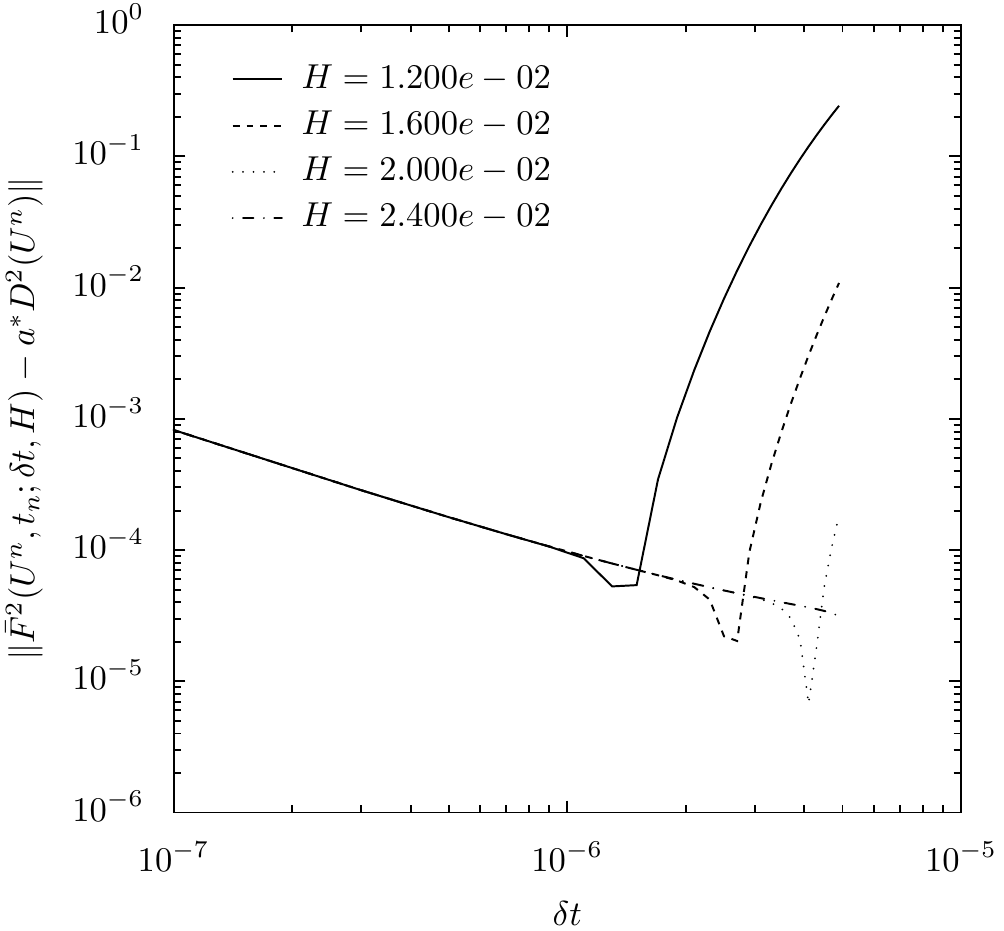}}
\caption{Error of the gap-tooth estimator $\bar{F}(U^n,t_n;\delta t,H)$ (using
the microscopic problem (\ref{eq:noflux_box}) with diffusion coefficient
(\ref{eq:diff_coeff}) in each box) with respect to the finite
difference time derivative $a^*D^2(U^n)$ on the same mesh.
\label{fig:err_ifv_buffers_noflux}} 
\end{figure}

\paragraph{Other boundary conditions}
In section \ref{subsec:consistency}, we studied the convergence of the
gap-tooth estimator both analytically and numerically in the case of Dirichlet
boundary conditions.  We will now show numerically that the results obtained in
that section do not depend crucially on the type of boundary conditions.
Consider again the diffusion problem (\ref{eq:model_eq_repeated}), with
the diffusion coefficient defined as in (\ref{eq:diff_coeff}), see also example
\ref{ex:3.6}.  We construct the gap-tooth time derivative estimator
$\bar{F}(U^n,t_n;\delta t,H)$ as outlined in section \ref{subsec:gap-tooth},
but now we use no-flux instead of Dirichlet boundary conditions.  In each box,
we then solve the following problem, 
\begin{equation}\label{eq:noflux_box}
\begin{split} 
\partial_t u_{\epsilon}(x,t)=\partial_x\left(a(x/\epsilon) \partial_x
u_{\epsilon}(x,t)\right),\\ 
\partial_x u_{\epsilon}(-H/2,t)=0,\qquad
\partial_x u_{\epsilon}(H/2,t)=0 
\end{split}
\end{equation} 
The concrete gap-tooth scheme that is used, as well as the corresponding finite
difference comparison scheme, are defined by the initialization
(\ref{eq:specific_initial}). Figure \ref{fig:err_ifv_buffers_noflux} shows the
error with respect to the finite difference time derivative $a^*D^2(U^n)$.
We see qualitatively the same behaviour as for Dirichlet boundary conditions.

In general, the choice of boundary conditions might influence the required
buffer size.  In the ideal case, where the boundary conditions are chosen to
correctly mimic the behaviour in the full domain, we can choose $H=h$.  Then
there is no buffer and the computational complexity is, in some sense, optimal.
For reaction-diffusion homogenization problems, this can be achieved by
constraining the averaged gradient around each box edge \cite{SamRooKevr03}.
In situations where the correct boundary conditions are not known, or prove
impossible to implement, one is forced to resort to the use of buffers.

\paragraph{Microscopic simulators}
It is possible that the microscopic model is not a partial differential
equation, but some microscopic simulator, e.g.~kinetic Monte Carlo or molecular
dynamics code.  In fact, this is the case where we expect our method to be most
useful.  In this case, the \emph{lifting} step, i.e.~the construction of box
initial conditions, becomes more involved.  In general, the microscopic model
will have many more degrees of freedom, the \emph{higher order moments} of the
evolving distribution.  These will quickly become slaved to the governing
moments (the ones where the lifting is conditioned upon), see
e.g.~\cite{KevrGearHymKevrRunTheo02,MakMarKevr02}. 
The crucial assumption in theorem \ref{thm:consistency} is that the solution
in each box evolves according to the macroscopic equation.  For a microscopic
simulation, this will usually mean that we need to construct an initial
condition in which, for example, a number of higher order moments are already
slaved to the governing moments (so-called \emph{mature} initial conditions).
To this end, it is possible to perform a constrained simulation before 
initialization to create such \emph{mature} initial conditions
\cite{Constraint03,HumKevr02}.  If this is not done, the resulting evolution
may be far from what is expected, see \cite{PvLLustKevr04} for an illustration
in the case of a lattice-Boltzmann model.

\section{Patch dynamics\label{sec:patch}}

Once a good gap-tooth time derivative estimator has been constructed, it can be
used as a method-of-lines spatial discretization in conjunction with any time
integration scheme.  Consider for concreteness the forward Euler scheme for
(\ref{eq:unknown}), given by
\begin{align}\label{eq:fe_comparison}
U^{n+1}&=U^n+\Delta t \; F(U^n,D^1(U^n),\ldots,D^d(U^n),t_n),\\
\intertext{which we will abbreviate as}
U^{n+1}&=U^n+\Delta t\; F(U^n,t_n)
\intertext{
and the corresponding patch dynamics scheme}
\label{eq:fe_patch}
\bar{U}^{n+1}&=\bar{U^n}+\Delta t \; \bar{F}^d(\bar{U}^n,t_n;\delta t,H),
\end{align}
where $\bar{F}^d(\bar{U}^n,t_n;\delta t,H)$ is defined as in (\ref{eq:der_est}).
Theorem \ref{thm:consistency} establishes the consistency of the gap-tooth
estimator.  In order to obtain convergence, we also need to prove stability.
For this purpose, we define the class $K$ of discrete functions with bounded
divided differences up to order $d$ on the numerical grid $(x_i,t_n)$,
$i=0,\ldots, N$; $t_n=n\Delta t$,
$n=0,\ldots,T/\Delta t$, as 
\begin{displaymath} 
K=\left\{\{U^n_i\}|\|D^{\alpha}_{\Delta x}U^n_i\| 
\le C_{\alpha} \text{ for } \alpha \le d, \; n\Delta t \le T \right\}, 
\end{displaymath} 
where $d$ is the highest spatial derivative present in equation
(\ref{eq:unknown}), $D^{\alpha}_{\Delta x}$ is the finite difference operator
of order $\alpha$ on a mesh of width $\Delta x$, and $C_{\alpha}$ is
independent of $\Delta t$ and $\Delta x$.  We can then make use of
\cite[Theorem~5.5]{EEng03} to state the following result.  (See also
\cite[Theorem~4.5]{SamRooKevr03} in the case of constrained gradient boundary
conditions.) 
\begin{thm} Consider the patch dynamics scheme (\ref{eq:fe_patch}) and the 
corresponding finite difference comparison scheme (\ref{eq:fe_comparison}). 
Assume that $\{\bar{U}^n\}$, $\{U^n\}$, $\{\hat{U}^n\} \in K$,
$U^0=\hat{U}^0=\bar{U}^0$ and the comparison scheme (\ref{eq:fe_comparison}) is
stable, then we have
\begin{displaymath} 
\|\bar{U}^n-u_0(x_i,t_n)\|\le C_1(\Delta x^k +\Delta t)+C_2
\max_{0\le k\le T/\Delta t} \|\bar{F}^d(\bar{U}^n,t_n;\delta t,H)-F(U^n,t_n)\|.
\end{displaymath} 
in the limit where 
\begin{displaymath}
\|\bar{F}^d(\bar{U}^n,t_n;\delta t,H)-F(U^n,t_n)\| \to 0.
\end{displaymath}
\end{thm} 
Thus the patch dynamics scheme is stable if the finite difference comparison
scheme is stable.  We note that, although this result is very general,
its applicability is limited due to the assumption that $\bar{U}^n \in K$, 
which has to be checked separately. Therefore, this result does not prevent the
patch dynamics scheme to become unstable, due to e.g.~an insufficient buffer
size $H$.  However, we can study the stability of the patch dynamics scheme
numerically by computing the eigenvalues of the time derivative estimator as a
function of $H$.  

\begin{figure}
\begin{center}
\includegraphics{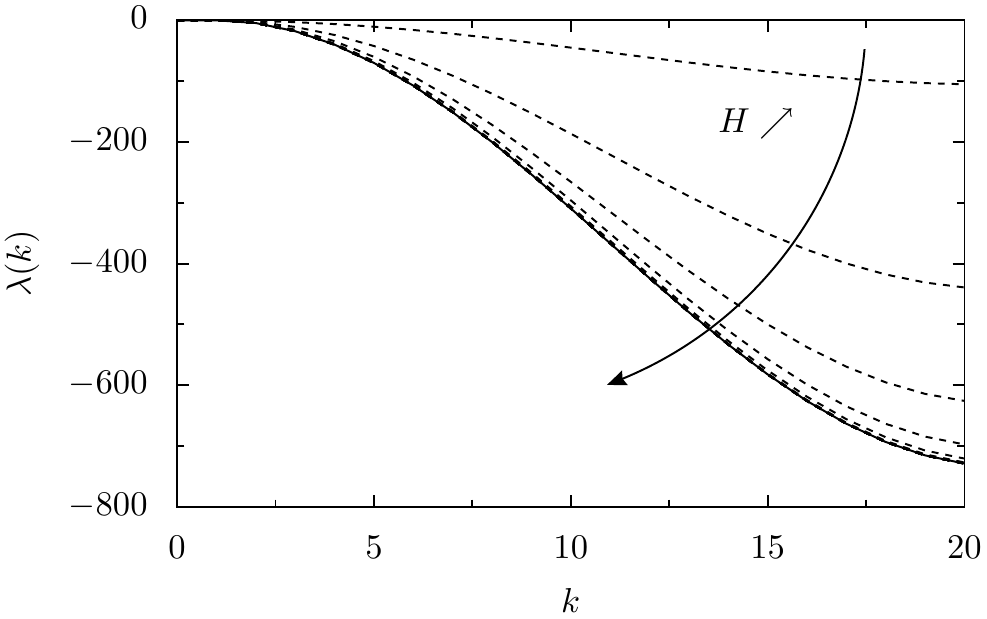}
\caption{\label{fig:spectrum_ifv_H} Spectrum of the estimator
$\bar{F}(U^n,t_n;\delta t,H)$ (dashed) for the model equation
(\ref{eq:model_eq_repeated}) with diffusion coefficient (\ref{eq:diff_coeff})
for $H=2\cdot 10^{-3},4\cdot 10^{-3},\ldots,2\cdot 10^{-2}$ and $\delta
t=5\cdot 10^{-6}$, and the eigenvalues (\ref{eq:ref_eigvals}) of $F(U^n,t_n)$
(solid).} 
\end{center}
\end{figure}

Consider the homogenization diffusion equation (\ref{eq:model_eq_repeated})
with the diffusion coefficient $a(x/\epsilon)$ given by (\ref{eq:diff_coeff}).
The homogenized equation is given by (\ref{eq:h_r}) with $a^*=0.45825686$.  In
this case, the time derivative operator $F(U^n,t_n)$ in the comparison scheme
(\ref{eq:fe_comparison}) has eigenvalues 
\begin{equation}\label{eq:ref_eigvals} 
\lambda_k=-\frac{4a^*}{\Delta x^2}\sin^2(\pi k \Delta x), 
\end{equation} 
which, using the forward Euler scheme as time-stepper, results in the stability
condition 
\begin{displaymath} 
\max_{k}\left|1+\lambda_k\Delta t\right|\le 1 \qquad \text{ or } \qquad 
\frac{\Delta t}{\Delta x^2}\le \frac{1}{2}a^*
\end{displaymath} 
It can easily be checked that the operator $\bar{F}(U^n,t_n;\delta t, H)$ is linear,
so we can interpret the evaluation of $\bar{F}(U^n,t_n;\delta t,H)$ as a
matrix-vector product.  We can therefore use any matrix-free linear algebra
technique to compute the eigenvalues of $\bar{F}(U^n,t_n;\delta t,H)$,
e.g.~Arnoldi. We choose to compute $\bar{F}(U^n,t_n;\delta t,H)$ and
$F(U^n,t_n)$ on the domain $[0,1]$ with Dirichlet boundary conditions, on a
mesh of width $\Delta x=0.05$ and with an inner box width of $h=2\cdot
10^{-3}$.  We choose $\delta t=5\cdot 10^{-6}$ and compute the eigenvalues
of $\bar{F}(U^n,t_n;\delta t,H)$ as a function of $H$.  The results are shown
in figure \ref{fig:spectrum_ifv_H}.
Two conclusions are apparent: since the most negative eigenvalue for
$\bar{F}(U^n,t_n;\delta t,H)$ is always smaller in absolute value than the
corresponding eigenvalue of $F(U^n,t_n)$ the patch dynamics scheme is always
stable if the comparison scheme is stable.  Moreover, we see that, with
increasing buffer size $H$, the eigenvalues of $\bar{F}(U^n,t_n;\delta t,H)$
approximate those of $F(U^n,t_n)$, which is an indication of consistency.

\section{Numerical results\label{sec:numeric}}

We will consider two example systems to illustrate the method.  The first
example is a system of two coupled reaction-diffusion equations, which models
CO oxidation on a heterogeneous catalytic surface.  Due to the reaction term,
the proof of theorem \ref{thm:consistency} is strictly speaking not valid, but
nevertheless the conclusions are the same.  The second example is the
Kuramoto--Sivashinsky equation.  This fourth-order non-linear parabolic equation
is widely used e.g.~in combustion modeling.  The patch dynamics scheme
with buffers also works in this case, showing the more general applicability of
the method.  All computations were performed in Python, making use of the SciPy
package \cite{SciPy} for scientific computing.

\subsection{Example 1: A nonlinear travelling wave in a heterogeneous 
excitable medium\label{sec:5.1}} 
Consider the following system of two coupled reaction-diffusion equations,
\begin{equation}\label{eq:co_eq}
\begin{split}
\partial_t u(x,t) &= \partial^2_x u(x,t)+\frac{1}{\delta}u(x,t)(1-u(x,t))
\left(u(x,t)-\frac{w(x,t)+b(x)}{a(x)}\right),\\
\partial_t w(x,t)&=g(u(x,t))-w(x,t),
\end{split}
\end{equation}
with
\begin{equation}\label{eq:co_nonlin}
g(u)=
\begin{cases}
0, & u < 1/3,\\
1-6.75\; u ( 1 - u )^2, & 1/3\le u < 1,\\
1, & u \ge 1.
\end{cases}
\end{equation}
This equation models the spatiotemporal dynamics of CO oxidation on
microstructured catalysts, which consist of, say, alternating stripes of two
different catalysts, such as platinum, Pt, and palladium, Pd, or platinum and
rhodium, Rh \cite{GrahKevrAsa94,BarBanKevr96,ShvartsSchutzImbKevr99}.  The goal
is to improve the average reactivity or selectivity by combining the catalytic
activities of the different metals, which are coupled through surface
diffusion.  In the above model, $u$ corresponds to the surface concentration of
CO, $w$ is a so-called surface reconstruction variable and $g(u)$ is an
experimentally fitted sigmoidal function.  Details can be found in
\cite{Keener00,Bar93}.  

In this model $a$ and $b$ and the time-scale ratio parameter $\delta$ are
physical parameters that incorporate the experimental conditions: partial
pressures of O$_2$ and CO in the gas phase, temperature, as well as kinetic
constants for the surface.  Here, we will study a domain of length $L=21$ with
a periodically varying medium: a striped surface that can be thought of as
consisting of equal amounts of Pt and Rh, with stripe width $\epsilon/2$.  The
medium is then defined by 
\begin{equation}\label{eq:co_par}
a(x)=0.84, \qquad b(x)=-0.025+0.725 \sin(2\pi x / \epsilon), \qquad
\delta = 0.025.
\end{equation} 
This particular choice of parameters is taken from \cite{RunTheoKevr02}, where
an effective bifurcation analysis for this model was presented.  
For these parameter values, the effective equation (given by
(\ref{eq:co_eq})-(\ref{eq:co_nonlin}) with 
\begin{equation}\label{eq:co_hpar}
a(x)=0.84, \qquad b(x)=-0.025, \qquad \delta = 0.025,
\end{equation}
supports travelling waves.  It was shown in \cite{RunTheoKevr02} that this
conclusion remains true for the given heterogeneity.  This was done by
computing the effective behaviour as the average of a large number of spatially
shifted realization of the wave.  Here, using the gap-tooth scheme, the
solution is spatially averaged inside each box, but the notion of effective
behaviour is identical.  We choose the small scale parameter $\epsilon=1\cdot
10^{-4}$.

The macroscopic comparison scheme for the effective equation
(\ref{eq:co_eq}-\ref{eq:co_nonlin})-(\ref{eq:co_hpar}) is defined as a standard
second order central difference discretization in space on a macroscopic mesh
of width $\Delta x=0.25$, combined with a forward Euler time-stepper. The
time-step is chosen as $\Delta t=1\cdot 10^{-2}$, which ensures stability. The
patch dynamics scheme for the detailed equation
(\ref{eq:co_eq}-\ref{eq:co_par}) is then obtained by using a gap-tooth
estimator for the time derivative using the initialization
(\ref{eq:specific_initial}) with the same forward Euler time-stepper.  

\begin{figure}
\begin{center}
\subfigure{\includegraphics[scale=0.7]{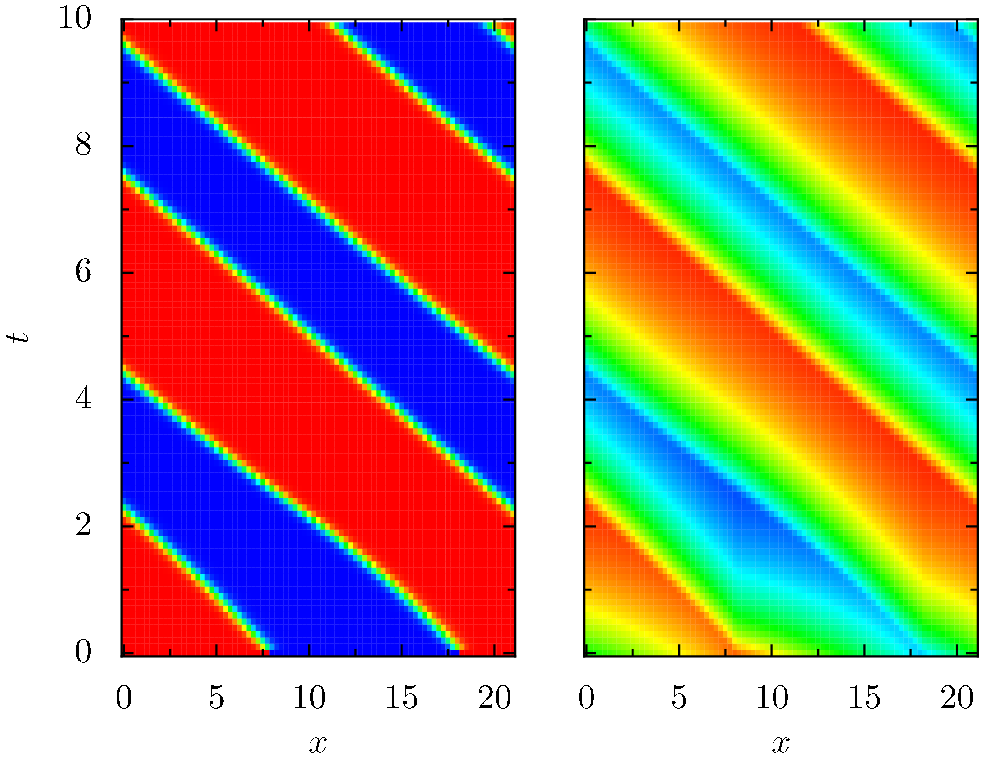}}
\subfigure{\includegraphics[scale=0.7]{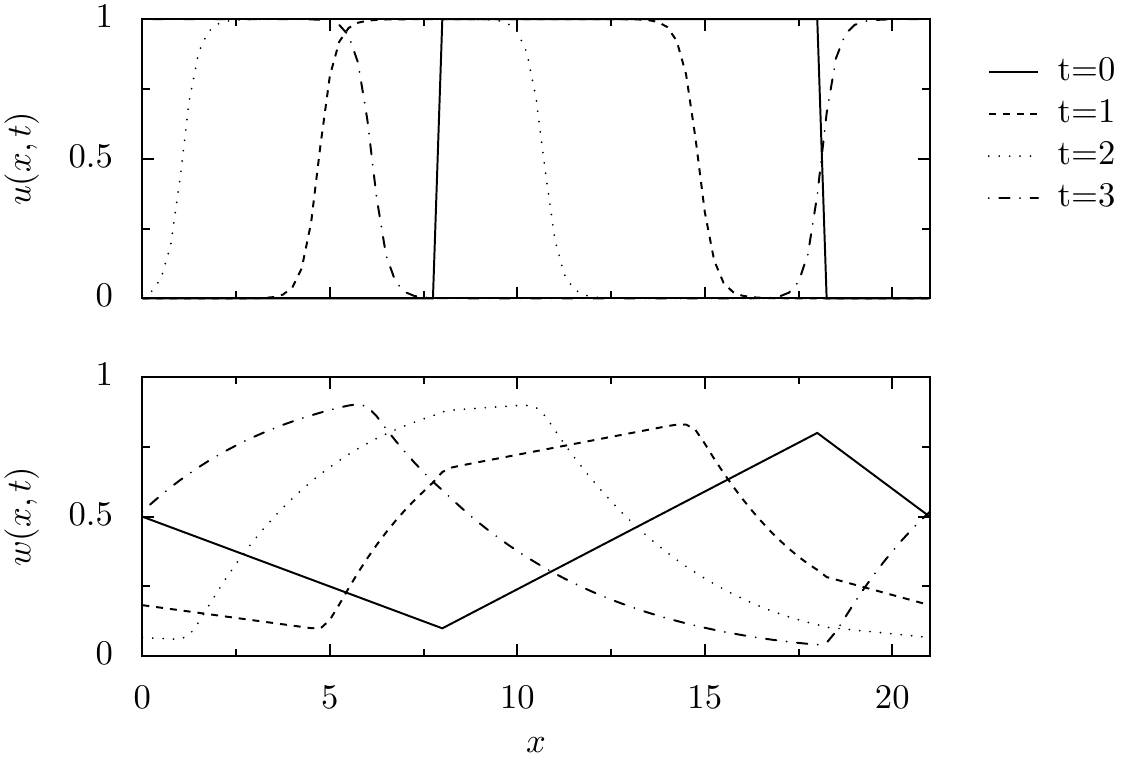}}
\caption{\label{fig:co_reaction_results}Left: solution of equation
(\ref{eq:co_eq}-\ref{eq:co_par}) using the patch dynamics scheme as a function
of space and time.  Colors indicate values (blue = 1, red = 0).  Right:
snapshots of the solution at certain moments in time, clearly showing the
approach to a travelling wave solution.} 
\end{center} 
\end{figure}

\begin{figure} 
\begin{center}
\includegraphics{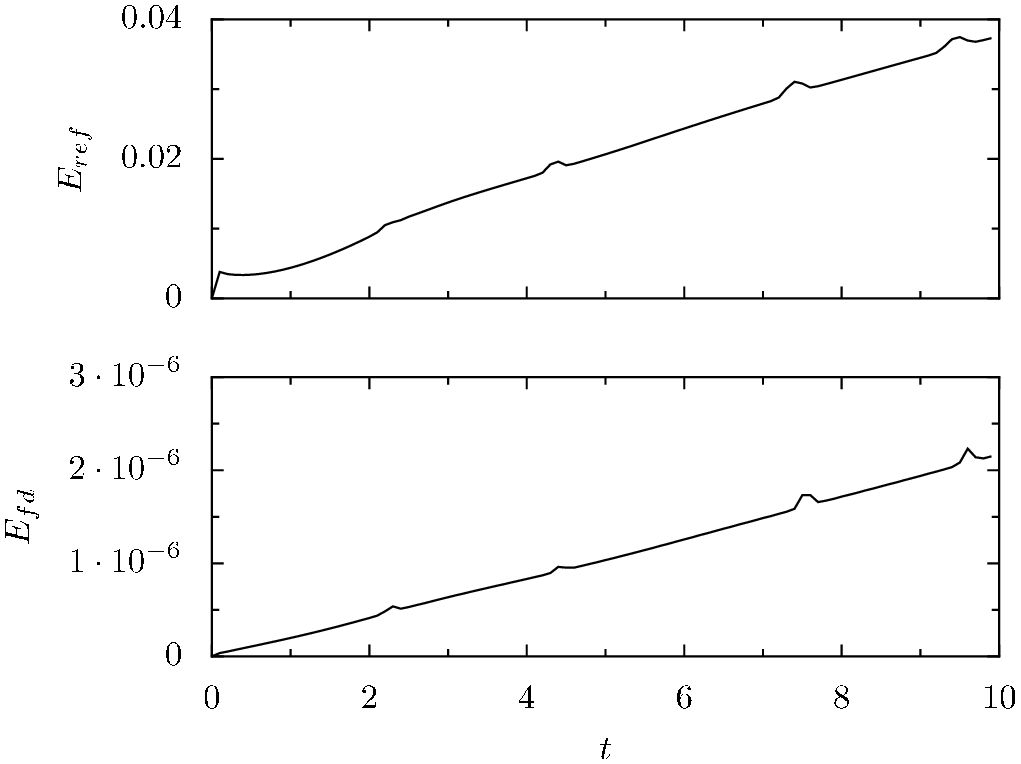} 
\caption{\label{fig:co_reaction_errors}
Error of a patch dynamics simulation for equation
(\ref{eq:co_eq}-\ref{eq:co_par}) with respect to the ``exact" solution of the
effective equation (top) and a finite difference comparison scheme (bottom).
The error is dominated by the error of the finite difference scheme.} 
\end{center} 
\end{figure}

\paragraph{Accuracy} We perform a numerical simulation for this model on the
domain $[0,L]$ using the patch dynamics scheme.  The gap-tooth
parameters are given by $h=5\cdot 10^{-4}$, $H=1.5\cdot 10^{-2}$ and $\delta
t=5\cdot 10^{-7}$.  Inside each box, we used a finite difference approximation
in space, with mesh width $\delta x=1\cdot 10^{-6}$ and \verb$lsode$ as
time-stepper.  The initial condition is given by 
\begin{equation*}
\begin{aligned} u(x,0)&=&
\begin{cases} 
1, & x \in [8,18]\\ 0, & 
\text{else}
\end{cases} 
\end{aligned}
\qquad 
\begin{aligned} 
w(x,0)&=& 
\begin{cases}0.5-0.05
x, & x \le 8,\\ 0.07 x -0.46 , & 8<x\le 18,\\ -0.1 x +2.6, & x > 18 .
\end{cases} 
\end{aligned} 
\end{equation*} 
The results are shown in figure \ref{fig:co_reaction_results}.  
We clearly see both the initial transient and the final travelling wave
solution.  For comparison purposes, the same computation was performed using
the finite difference comparison scheme for the effective equation.  We also
computed an ``exact'' solution for the effective equation using a much finer
grid ($\Delta x=5\cdot 10^{-3}$ and $\Delta t=1\cdot 10^{-5}$.  Figure
\ref{fig:co_reaction_errors} shows the errors of the patch dynamics simulation
with respect to the finite difference simulation of the effective equation and
the ``exact'' solution, respectively.  We clearly see that the patch dynamics
scheme is a very good approximation of the finite difference scheme, and the
error with respect to the exact solution is dominated by the error of the
finite difference scheme.  

\paragraph{Efficiency} Time integration using the patch dynamics scheme is more
efficient than a complete simulation using the microscopic model, since the
microscopic model is used only in small portions of the space-time domain (the
patches).  An obvious (but not always correct) way to study the efficiency is
to compare the size of the total space-time domain with the size of the
patches.  In this example, the simulations are only performed in 6\% of the
spatial domain.  Of course, when it is possible to apply physically correct
boundary conditions around the inner box, the buffer boxes are not necessary,
and the boxes would only cover 0.2 \% of the space domain.  For
reaction-diffusion homogenization problems, we showed that buffer boxes are not
required when we constrain the average gradient at the box boundary
\cite{SamRooKevr03}.  The gain in the spatial dimension is determined by the
separation in spatial scales.  It can be large when the macroscopic solution is
smooth (few macroscopic mesh points are needed) and propagation of boundary
artefacts is slow (small buffer box is sufficient). Note that in higher spatial
dimensions, this gain can be even more spectacular.

The gain in the temporal dimension can be determined similarly.  In the example
of section \ref{sec:5.1}, the gap-tooth step was chosen as $\delta t=5\cdot
10^{-7}$, whereas for macroscopic time integration, the forward Euler scheme
was used with $\Delta t=1\cdot 10^{-2}$.  Therefore, in the temporal dimension,
we gain a factor of $2\cdot 10^{5}$.  In more realistic applications, when the
microscopic model is not a partial differential equation, we expect this gain
to be smaller, since additional computational effort will be required to remove
the errors that were introduced during the lifting step, e.g. in the form of
constrained simulation \cite{Constraint03}.

\subsection{Example 2: Kuramoto--Sivashinsky equation\label{sec:5.2}}

Consider the Kuramoto--Sivashinsky equation
\begin{equation}\label{eq:kur}
\partial_t u(x,t)=-\nu \partial_x^4 u(x,t) - 
\partial_x^2 u(x,t)-u(x,t)\partial_x u(x,t),\qquad x \in [0,2\pi],
\end{equation}
with periodic boundary conditions. This equation is frequently used in the
modelling of combustion and thin film flow.  For the parameter value
$\nu=4/15$, it has been shown that the equation supports travelling wave
solutions, see e.g.~\cite{KevrNicoScov90}.  For the purpose of this example,
both the microscopic and the macroscopic model are given by (\ref{eq:kur}).  

To obtain the macroscopic comparison scheme, we discretize the second and 
fourth order spatial derivatives using second order central differences, on a
macroscopic mesh of width $\Delta x = 0.05\pi$, combined with a forward Euler
time integrator with time-step $\Delta t = 1\cdot 10^{-5}$.  This small 
macroscopic time-step arises due to the stiffness of the effective equation.
We can accelerate time-stepping by wrapping a so-called \emph{projective
integration method} around the forward Euler scheme \cite{GearKevr01}.  This
scheme works as follows.  First, we perform a number of forward Euler steps,
\begin{align*}
U^{k+1,N}&=U^{k,N}+\Delta t \; F(U^{k,N},t_n),\\
\intertext{
where, for consistency, $U^{0,N}=U^N$, followed by a large extrapolation step}
U^{N+1}&=(M+1)\;U^{k+1,N}-M\; U^{k,N}, \qquad M>k.
\end{align*}
Here, $U^{k,N}\approx U(N\;(M+k+1)\;\Delta
t+k\Delta t)$.  The parameters $k$ and $M$ determine the stability region
of the resulting time-stepper.  An analysis of these methods is given in
\cite{GearKevr01}. It can be checked that, for this equation, choosing $k=2$
and $M=7$ results in a stable time-stepping scheme.

The patch dynamics scheme is constructed by replacing the time derivative
$F(U^{k,N},t_n)$ by a gap-tooth estimator $\bar{F}^4(\bar{U}^{k,N},t_n;\delta
t,H)$, obtained by the initialization (\ref{eq:local_taylor}), where we choose
the order of the Taylor expansion to be $d=4$. The coefficients $D_k^i$, $k>0$
are determined by the macroscopic comparison scheme.  Inside each box, equation
(\ref{eq:kur}) is solved, on a mesh of width $\delta x =1\cdot 10^{-5}$,
subject to Dirichlet and no-flux boundary conditions, using \verb#lsode# as
time-stepper. We fixed the box width $h=1\cdot 10^{-3}$.

\begin{figure}
\begin{center}
\subfigure{
\includegraphics[scale=0.7]{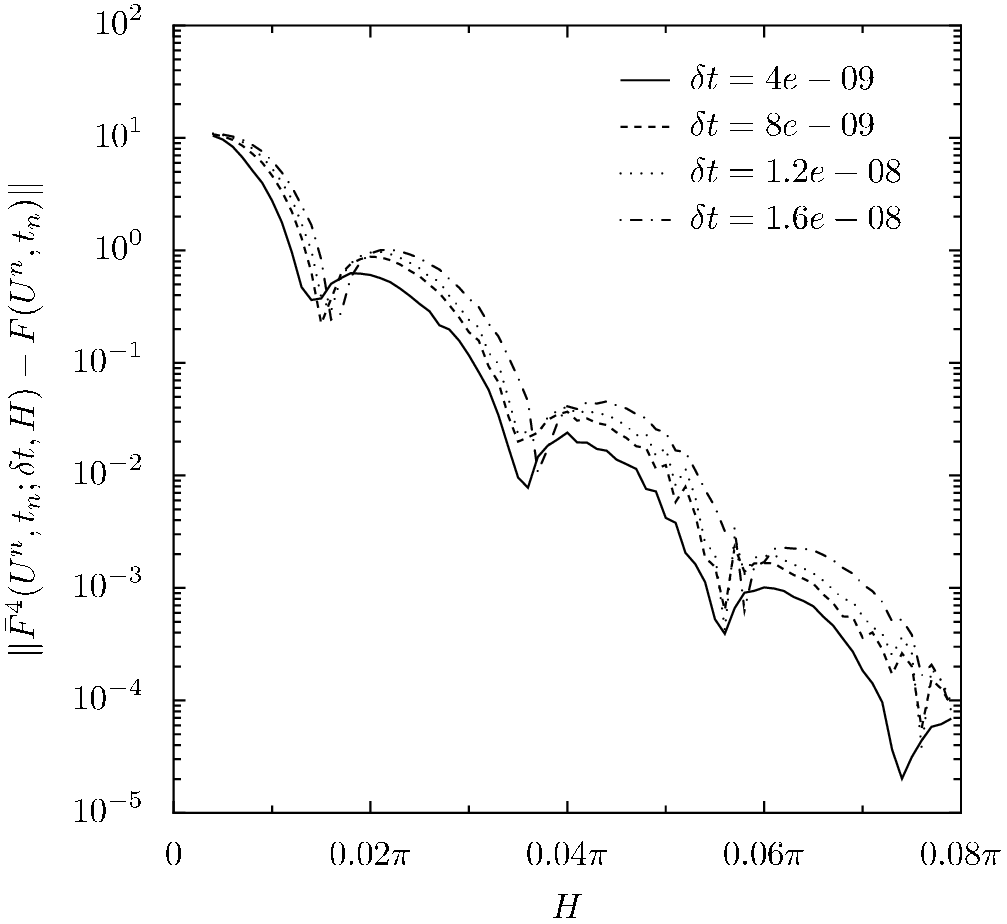}}
\subfigure{
\includegraphics[scale=0.7]{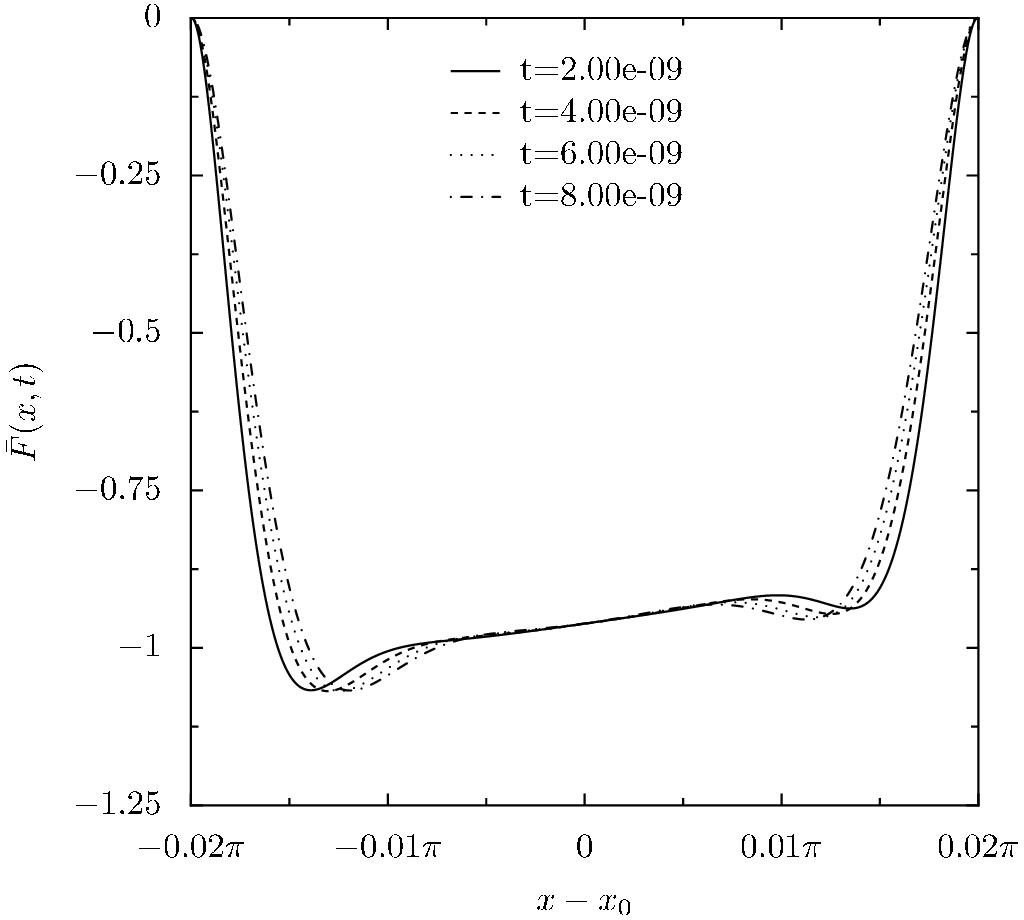}}
\caption{\label{fig:cons_kur}Left: Error of the gap-tooth estimator 
$\bar{F}(\bar{U}^{0,0},t_n;\delta t,H)$ with respect to the finite difference 
time derivative $F(U^{0,0},t_n)$.  Right:  The function (\ref{eq:est_ifv_x}) as
a function of $x$ for a number of values of time.  We clearly see how
the estimate gets affected by the boundary conditions.  }
\end{center}
\end{figure}

\paragraph{ Consistency and efficiency }  
Because of the fourth order term, theorem \ref{thm:consistency} is not proven.
Therefore, we numerically check the consistency of the estimator, by computing
the gap-tooth estimator $\bar{F}^4(\bar{U}^{k,N},t_n;\delta t,H)$ as a function
of $H$ for a range of values for $\delta t$, and comparing the resulting
estimate with the time derivative of the comparison scheme.  As an initial
condition, we choose $u^0(x)=\sin(2\pi x)$.  The results are shown in figure
\ref{fig:cons_kur} (left).  
We see qualitatively the same behaviour as in section \ref{subsec:consistency}
for diffusion problems.  There are two main differences.  First, in this case
the convergence is no longer monotonic, which explains the sharp peaks in the
error curves.  Also, because boundary artefacts travel inwards much faster, the
gain will be much smaller.  Indeed, the figure suggests that a good compromise
between accuracy and efficiency would be to choose $\delta t = 4\cdot 10^{-9}$
and $H=3\pi\cdot 10^{-2}$.  The reason for this behaviour is that the
macroscopic equation contains reasonably fast time scales.  Note that this is
also the reason why the finite difference comparison scheme is forced to take
small time-steps.   

For this choice of the parameters, the computations have to be performed in
60\% of the spatial domain. However, for a forward Euler step, we only need to
simulate in 1/25000 of the time-domain.  Using the projective integration
scheme therefore gives us a total gain factor of about 80000 in time.  Again,
we note that in real applications, this spectacular gain will partly be
compensated by the additional computational effort that is required to create
appropriate initial conditions.  

We can draw two main conclusions.  The scheme allows to simulate higher order
macroscopic equations, and the gain in the space domain is heavily
dependent on the separation of scales in the macroscopic equation.

\begin{figure}
\begin{center}
\subfigure{\includegraphics[scale=0.62]{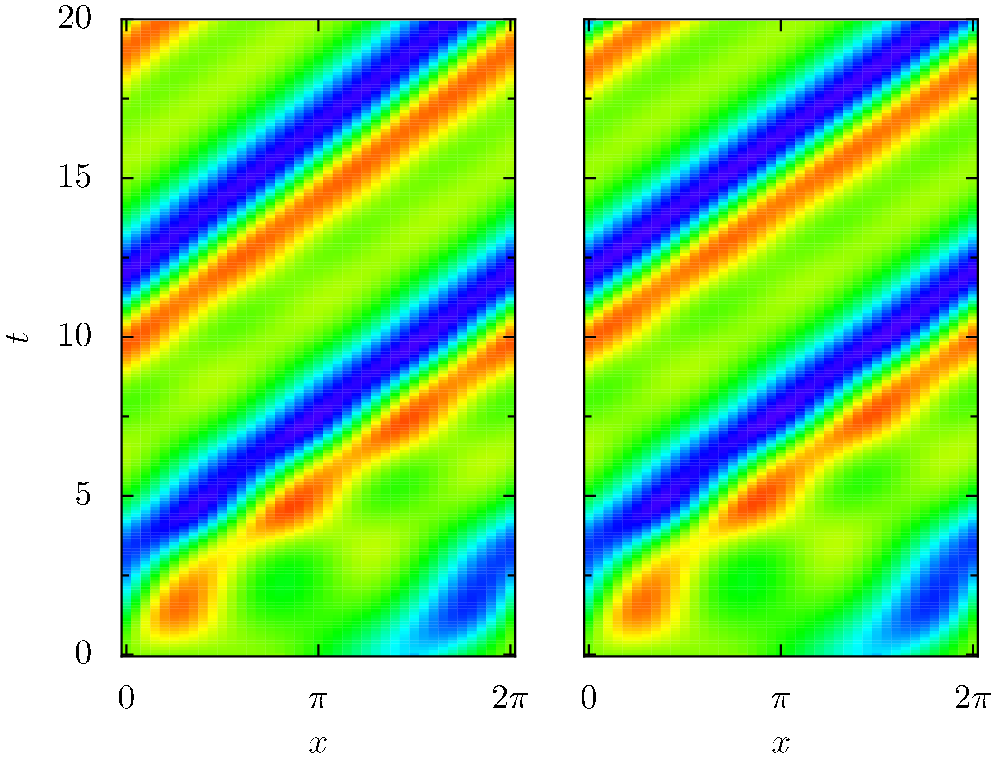}}
\subfigure{\includegraphics[scale=0.62]{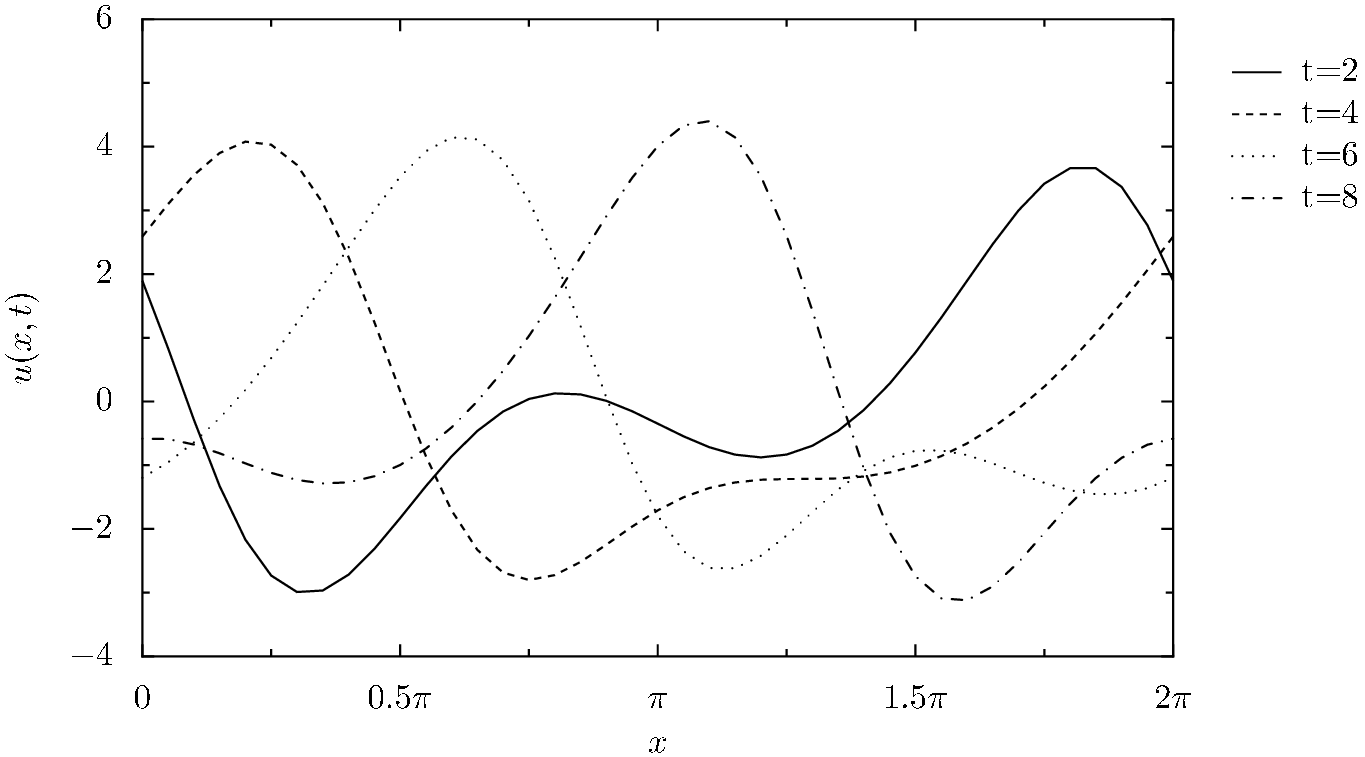}}
\caption{\label{fig:kur_reaction_results}Left: solution of equation
(\ref{eq:kur}) using the patch dynamics scheme as a function of space and time.
Colors indicate values (blue = 4, red = -4).  Right: snapshots of the solution
at certain moments in time, clearly showing the approach to a travelling wave
solution.} 
\end{center} 
\end{figure}

\begin{figure} 
\begin{center}
\includegraphics{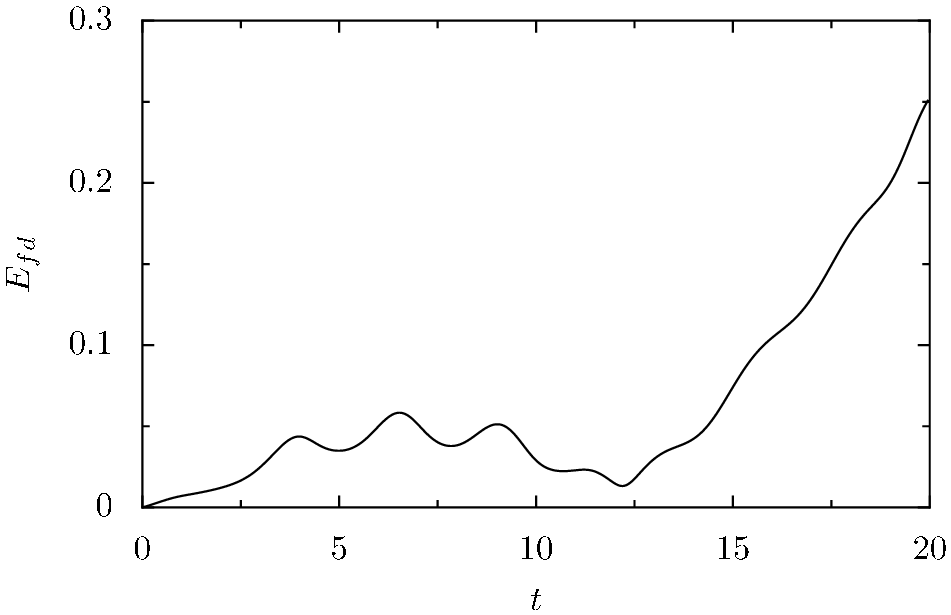} 
\caption{\label{fig:kur_reaction_errors}
Error of a patch dynamics simulation for equation (\ref{eq:kur}) with respect
to the a finite difference comparison scheme for the effective equation. We see
that this error grows monotonic once the travelling wave has been reached, due
to a slight difference in propagation speed.} 
\end{center} 
\end{figure}

\paragraph{Accuracy} We perform a numerical simulation for this model on the
domain $[0,2\pi]$ using the patch dynamics scheme.  The gap-tooth
parameters are given by $h=1\cdot 10^{-3}$, $H=3\pi\cdot 10^{-2}$ and $\delta
t=4\cdot 10^{-9}$.  Inside each box, we used a finite difference approximation
in space, with mesh width $\delta x=1\cdot 10^{-5}$ and \verb$lsode$ as
time-stepper.  The initial condition is given by 
\begin{equation*}
u(x,0)=
\begin{cases} 
-1, & x \in [0,0.8\pi],\\ 
-1+5(x-0.8\pi), & x\in [0.8\pi,1.5\pi], \\ 
2.5-7(x-1.5\pi), & x\in [1.5\pi,2\pi],
\end{cases} 
\end{equation*} 
The results are shown in figure \ref{fig:kur_reaction_results}.  
We clearly see both the initial transient and the final travelling wave
solution.  For comparison purposes, the same computation was performed using
the finite difference comparison scheme for the effective equation.  Figure
\ref{fig:kur_reaction_errors} shows the errors of the patch dynamics simulation
with respect to the finite difference simulation.  
We see that during the transient phase the error oscillates somewhat, but
once the travelling wave is steady the error increases linearly, due to
a difference in the approximated propagation speed.  Note that the error is
significantly larger than for example 5.1, due to the fact that the estimator
is less accurate, but also because the macroscopic time-step is much smaller,
resulting in a larger number of estimations.

\section{Conclusions\label{sec:conclusions}}

We described the patch dynamics scheme for multi-scale problems.  This scheme
approximates an unavailable \emph{effective} equation over macroscopic
time and length scales, when only a microscopic evolution law is given; it only
uses appropriately initialized simulations of the microscopic model over small
subsets (patches) of the space-time domain.  Because it is often not possible
to impose macroscopically inspired boundary conditions on a microscopic
simulation, we propose to use buffer regions around the patches, which
temporarily shield the internal region of the patches from boundary artefacts.

We analytically derived an error estimate for a model homogenization problem
with Dirichlet boundary conditions.  The numerical results show that the
algorithm is more widely applicable.  We showed the scheme is capable of giving
good approximations for reaction-diffusion systems, as well as for fourth order
PDEs, such as the Kuramoto--Sivashinsky equation.  As such, these results are
far more general than those of \cite{SamRooKevr03}, which are restricted to
reaction-diffusion problems due to the special choice of boundary conditions.

We emphasize that, although analyzed for homogenization problems, the real
advantage for the methods presented here lies in their applicability for
microscopic models that are not PDEs, such as kinetic Monte Carlo, or molecular
dynamics.  Experiments in this direction are currently being pursued actively.

\section*{Acknowledgments} Giovanni Samaey is a Research Assistant of the Fund
for Scientific Research - Flanders.  This work has been partially supported by
grant IUAP/V/22 and by the Fund of Scientific Research through Research Project
G.0130.03 (GS, DR), and an NSF/ITR grant and AFOSR Dynamics and Control, Dr. S.
Heise (IGK).

\end{document}